\documentclass[11pt]{article}

\usepackage[top=1.0in,right=1.0in,left=1.0in,bottom=1.75in]{geometry}
\usepackage{amssymb,amsbsy}
\usepackage{amsmath}
\usepackage{amsthm}
\usepackage{graphics,graphicx}
\usepackage[T1]{fontenc}
\usepackage{color}
\usepackage{subfigure}
\usepackage[affil-it,auth-sc]{authblk}
\usepackage{stmaryrd}

\usepackage[]{jabbrv}

\usepackage[numbers,sort&compress]{natbib}

\usepackage{soul}
\definecolor{lightyellow}{rgb}{1.0, 1.0, 0.88}
\definecolor{lightmauve}{rgb}{0.86, 0.82, 1.0}
\sethlcolor{lightmauve}

\newcommand{\beq}{\begin{equation}}
\newcommand{\eeq}{\end{equation}}

\newcommand{\beqar}{\begin{eqnarray}}
\newcommand{\eeqar}{\end{eqnarray}}
\newcommand{\bit}{\begin{itemize}}
\newcommand{\eit}{\end{itemize}}
\newcommand{\benum}{\begin{enumerate}}
\newcommand{\eenum}{\end{enumerate}}
\newcommand{\barr}{\begin{array}}
\newcommand{\earr}{\end{array}}

\newcommand\eq[1]{(\ref{#1})}

\def\XXint#1#2#3{{\setbox0=\hbox{$#1{#2#3}{\int}$}
   \vcenter{\hbox{$#2#3$}}\kern-.5\wd0}}

\def\b0{\mbox{\boldmath $0$}}

\def\bk{\mbox{\boldmath $k$}}

\def\bC{\mbox{\boldmath $C$}}

\def\f0{\ensuremath{\mathbb{O}}}

\newcommand{\Go}{\omega}

\newcommand{\GG}{\Gamma}

\newcommand{\bmA}{\mbox{\boldmath $\mathcal{A}$}}

\graphicspath{{./}{./figures/}}

\title{Dispersion degeneracies and standing modes in flexural waves \\ supported by Rayleigh beam structures}

\author[1]{A. Piccolroaz\footnote{Corresponding author: e-mail: roaz@ing.unitn.it; phone: +39\,0461\,282583.}}
\author[2]{A.B. Movchan}
\author[1]{L. Cabras}

\affil[1]{Dipartimento di Ingegneria Meccanica e Strutturale, Universit\`a di Trento, Italy}
\affil[2]{Department of Mathematical Sciences, University of Liverpool, U.K.}

\date{}

\begin{document}

\maketitle

\begin{abstract}
\noindent
The paper presents a novel analysis of Floquet-Bloch flexural waves in a periodic lattice-like structure consisting of flexural beam ligaments. A special feature of this structure is in the presence of the rotational inertia, which is commonly neglected in  conventional models of the Euler-Bernoulli type.
The dispersion properties of the Rayleigh beam structure with rotational inertia include degeneracies linked to Dirac cones on the dispersion diagrams as well as directional anisotropy and special refraction properties.
Steering of Dirac cones is described for rectangular flexural structures with a rotational inertia.
Numerical examples for a forced network of Rayleigh and Euler-Bernoulli beams illustrate directional localisation, negative refraction, localisation at an interface and neutrality for propagating plane waves across a structured interface for a frequency range corresponding to a Dirac cone.
\end{abstract}

{\it Keywords: Rayleigh beam; Rotational inertia; Dispersive waves; Metamaterial}


%

\section{Introduction}
\label{sec01}

Dispersion of Floquet-Bloch waves in periodic flexural systems is a topic of high importance, that attracts attention of experts across a wide range of fields. In particular, we would like to refer to the work by Slepyan \cite{Slepyan_2002} and Slepyan and Ryvkin \cite{Slepyan_Ryvkin_2010} who studied Floquet-Bloch waves in conjunction with the problems of dynamic fracture of elastic lattices. For periodic structures of elastic beams, the explicit solutions were also obtained and analysed by Heckl \cite{Heckl_2002}, Brun {\em et al.} \cite{Brun_2013} with the emphasis on analytical modelling of Floquet-Bloch waves and propagation of damage.
The papers by Bigoni and Movchan \cite{Bigoni_2002}, and Gei {\em et al.}\cite{Gei2009} have dealt with transmission problems for solids with structured interfaces and have presented an asymptotic approach to a derivation of the model and analysis of Floquet-Bloch waves in pre-stressed periodic layers on an elastic half-space, as well as control of band gaps on the dispersion diagram by applying a prestress to the elastic system.

In the recent paper by Piccolroaz and Movchan \cite{PM_2014}, a comparative analysis was presented for Floquet-Bloch waves in structured Rayleigh and Euler beams subjected to prestress and the action from an elastic foundation. There was also a formal connection made to formulations of the constrained Cosserat approach in antiplane elasticity, or a couple stress analysis. In particular, it has been
demonstrated in \cite{PM_2014} that the waves in Rayleigh beams become non-dispersive at high frequencies, whereas at low frequency the waves for both Rayleigh and Euler beams are similar. Effects of prestress discussed in \cite{PM_2014} have been presented in closed analytical form, which was also followed by description of defects based on dynamic Green's functions.
Of course, one-dimensional studies \cite{PM_2014} did not allow for the dynamic anisotropy, and we would like  to study here the anisotropy in the dynamic response of the Euler and Rayleigh two-dimensional flexural systems.
Although it is not the purpose of the present paper to discuss possible experimental or manufacturing implementation of Rayleigh beams, we would like to refer to the earlier paper \cite{PM_2014} which includes the discussion of possible implementations by embedding rotational resonators distributed along the beam (``hedgehog-like'' structure).
It is interesting that the governing equation for the flexural displacement brings an additional term containing the second-order spatial derivative of the displacement, which is also proportional to the square of the frequency. This term may play together with the pre-stress force leading to a dynamic buckling and novel dispersion properties for periodically structured Rayleigh beams, as it has been outlined in \cite{PM_2014}. Classification and applications of non-classical beam theories are discussed in detail in \cite{Timoshenko1990, HAN1999935, Li20111677}. A particular emphasis is made here on the difference in the dynamic response between the structures of the Euler-Bernoulli and of the Rayleigh beams in the intermediate frequency range. Indeed in the quasi-static regime the response of these two classes of structures to external loading becomes similar. Special attention is given to a dynamic anisotropy, which is linked to the design of hyperbolic metamaterial structures.

The purpose of the present study is to address the challenge of analytical description of a dynamically anisotropic response of doubly periodic structures of beams in the cases of Euler-Bernoulli beams and the Rayleigh beams.
Specifically we are looking for the frequency regimes where the difference between the Rayleigh beams and the  Euler-Bernoulli beams becomes significant, and we identify regimes for which the Rayleigh beams systems exhibit negative refraction if analysed as a cluster subjected to an incident wave.

Dispersion surfaces and the isofrequency contours (or slowness contours) contain significant information about standing wave regimes, band gaps and dynamic anisotropy.
We construct the dispersion surfaces and give a comparative analysis for both Rayleigh beams as well as the Euler-Bernoulli beams. Special interest is deserved by the so-called Dirac points representing multiple roots of the dispersion equation, where the dispersion surfaces may become non-smooth.

\section{Formulation of the problem for Floquet-Bloch waves in Rayleigh beam lattices}
\label{sec02}

\begin{figure}[!htb]
\centering
\includegraphics[width=150mm]{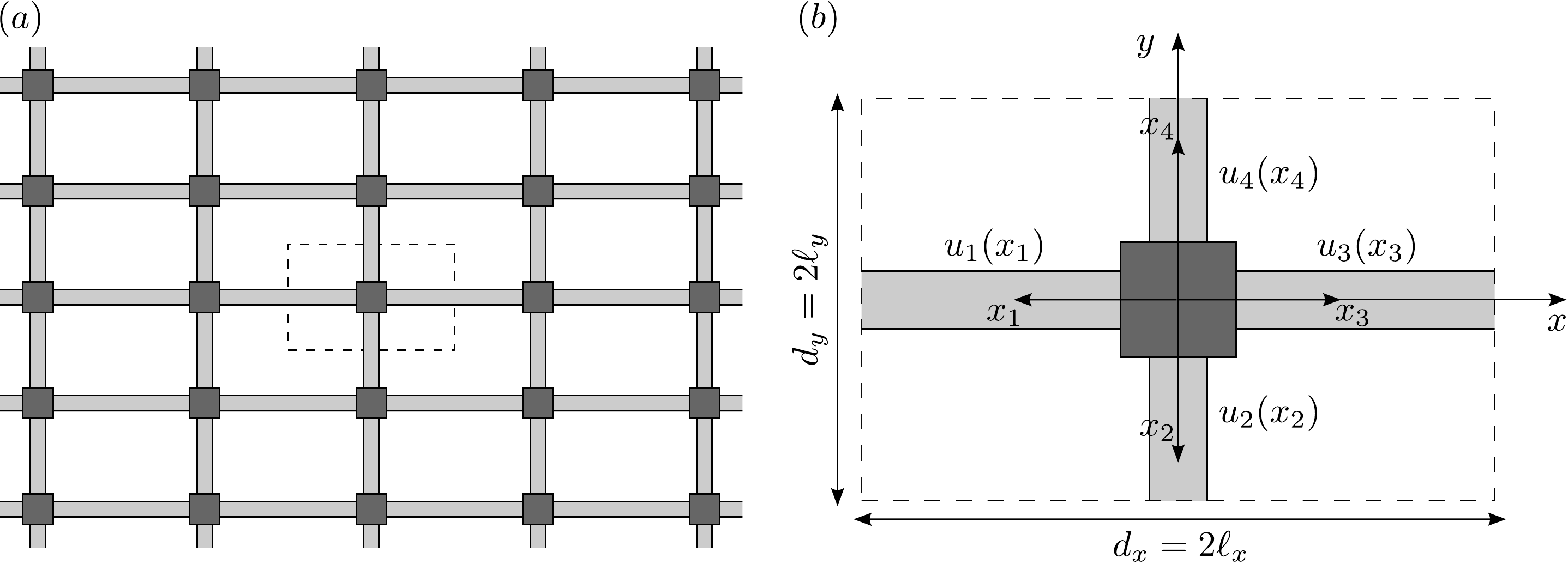}
\caption{\footnotesize (a) An infinite rectangular periodic network of Rayleigh beams with the elementary cell shown as the dashed line rectangle. (b) The elementary cell showing the local coordinates for each beam element.}
\label{fig01}
\end{figure}

We consider an infinite rectangular periodic network of Rayleigh beams, as shown in Fig.~\ref{fig01}, and assume that the beams can deflect in the out-of-plane direction.

The governing equation for time-harmonic flexural waves in a Rayleigh beam is
\begin{equation}
\label{eq:gov}
EI u''''(x) - (P - \rho I \omega^2) u'' + (\beta - \rho A \omega^2) u = 0,
\end{equation}
where $E$ is the Young modulus, $\beta$ is the stiffness of a Winkler type elastic foundation, $P$ the prestress, $\rho$ the mass density, $A$ the area of the cross-section, and $I$ the area moment of inertia of the cross-section.

The internal bending moment $M$ and the internal shear force $V$ are given by
\begin{equation}
M(x) = -EI u''(x), \quad V(x) = -EI u'''(x) + (P - \rho I \omega^2) u'(x),
\end{equation}
respectively.

The solution of (\ref{eq:gov}) is sought in the form
\begin{equation}
u(x) = C e^{i \kappa x},
\end{equation}
which yields the characteristic roots in the form
\begin{equation}
\kappa_{1,2,3,4} = \pm \frac{1}{r} \sqrt{-\frac{\overline{P} - R\omega^2}{2} \pm \sqrt{\frac{(\overline{P} - R\omega^2)^2}{4} + R\omega^2 - \overline{B}}},
\end{equation}
where
\begin{equation}
r = \sqrt{\frac{I}{A}}, \quad
\overline{P} = \frac{Pr^2}{EI} = \frac{P}{EA}, \quad
\overline{B} = \frac{\beta r^4}{EI} = \frac{\beta I}{EA^2}, \quad
R = \frac{\rho r^2}{E} = \frac{\rho I}{EA}.
\end{equation}

Assuming that there are no double roots, the general solution is then given by the linear combination
\begin{equation}
u(x) = \sum_{q=1}^{4} C_q e^{i \kappa_q x},
\end{equation}
In order to set up the boundary value problem for the beams within the unit cell $[-\ell_x,\ell_x]\times[-\ell_y,\ell_y]$, it is convenient to introduce local coordinates for each beam, as shown in Fig.~\ref{fig01},
\begin{equation}
u_1(x_1) = u(-x), \quad u_2(x_2) = u(-y), \quad u_3(x_3) = u(x), \quad u_4(x_4) = u(y),
\end{equation}
and write the general solution for each beam as follows
\begin{equation}
u_p(x_p) = \sum_{q=1}^{4} C_{pq} e^{i \kappa_{pq} x_q}, \quad p=1,2,3,4.
\label{Cpq}
\end{equation}
There are 16 undetermined constants, $C_{pq}$, $p=1,\cdots,4$, $q=1,\cdots,4$, which can be found by the 8 Floquet-Bloch conditions at the boundary of the unit cell supplemented by the 8 junction conditions at the central node of the unit cell.

In particular, for the horizontal beams, the quasi-periodic conditions at $x = -\ell_x$ and $x = \ell_x$ hold
\begin{equation}
u_3(\ell_x) = u_1(\ell_x) e^{i 2\ell_x k_x},
\end{equation}
\begin{equation}
u_3'(\ell_x) = -u_1'(\ell_x) e^{i 2\ell_x k_x},
\end{equation}
\begin{equation}
- EI u_3''(\ell_x) = - EI u_1''(\ell_x) e^{i 2\ell_x k_x},
\end{equation}
\begin{equation}
-EI u_3'''(\ell_x) + (P_x - \rho I \omega^2) u_3'(\ell_x) =
- \big[ -EI u_1'''(\ell_x) + (P_x - \rho I \omega^2) u_1'(\ell_x) \big] e^{i 2\ell_x k_x},
\end{equation}
which prescribe the Floquet-Bloch shift across the unit cell along the horizontal direction for flexural displacement, rotation, internal moment and internal shear force. Analogous quasi-periodic boundary conditions apply for the vertical beams at $y = -\ell_y$ and $y = \ell_y$,
\begin{equation}
\label{eq:cond1}
u_4(\ell_y) = u_2(\ell_y) e^{i 2\ell_y k_y},
\end{equation}
\begin{equation}
u_4'(\ell_y) = -u_2'(\ell_y) e^{i 2\ell_y k_y},
\end{equation}
\begin{equation}
- EI u_4''(\ell_y) = - EI u_2''(\ell_y) e^{i 2\ell_y k_y},
\end{equation}
\begin{equation}
-EI u_4'''(\ell_y) + (P_y - \rho I \omega^2) u_4'(\ell_y) =
- \big[ -EI u_2'''(\ell_y) + (P_y - \rho I \omega^2) u_2'(\ell_y) \big] e^{i 2\ell_y k_y}.
\end{equation}
The junction conditions at the central node require continuity of flexural displacements
\begin{equation}
u_3(0) = u_1(0),
\end{equation}
\begin{equation}
u_2(0) = u_1(0),
\end{equation}
\begin{equation}
u_4(0) = u_1(0),
\end{equation}
continuity of rotations
\begin{equation}
u_3'(0) = -u_1'(0),
\end{equation}
\begin{equation}
u_4'(0) = -u_2'(0),
\end{equation}
and, finally, equation of motion for the central node
\begin{multline}
-\big[ -EI u_1'''(0) + (P_x - \rho I \omega^2) u_1'(0) \big]
-\big[ -EI u_3'''(0) + (P_x - \rho I \omega^2) u_3'(0) \big] \\
-\big[ -EI u_2'''(0) + (P_y - \rho I \omega^2) u_2'(0) \big]
-\big[ -EI u_4'''(0) + (P_y - \rho I \omega^2) u_4'(0) \big] = 
M_0 \omega^2 u_1(0),
\end{multline}
\begin{equation}
-EI u_3''(0) + EI u_1''(0) = I_{0y} \omega^2 u_3'(0),
\end{equation}
\begin{equation}
\label{eq:cond16}
-EI u_4''(0) + EI u_2''(0) = I_{0x} \omega^2 u_4'(0),
\end{equation}
where $M_0, I_{0x}, I_{0y}$ are the mass and moments of inertia of the central node.

Eqs. (\ref{eq:cond1})--(\ref{eq:cond16}) provide a homogeneous linear system for the 16 unknown constants so that the vanishing of the determinant of the associated matrix yields the dispersion equation for the Rayleigh beam lattice.

\section{Lower-dimensional model, dispersion equation}
\label{sec03}

Analysis of Floquet-Bloch waves is very helpful in identifying the dynamic response of structured solids.
As in the classical homogenisation theories (see, for example, the monographs \cite{BLP1978, SP1980, Jikov1994}), the overall static response of structured solids can be achieved by a suitably designed averaging procedure used in the studies of partial differential equations with rapidly oscillating coefficients, many of these theories fail in the case of dynamics, where elastic waves propagate through the structure and create dynamic vibration modes. In the latter case, the theory of Floquet-Bloch systems, which refers to an infinite periodic structure and in which the quasi-periodicity boundary conditions (also known as the Floquet-Bloch conditions) are set on the boundary, is efficiently used as the means of describing the dynamic response of structured solids (see, for example, \cite{Physics_2013}).
The high-frequency homogenisation (HFH) has also been developed in \cite{Craster_2014} in order to identify anisotropy and spatial localisation for states adjacent to standing waves of zero group velocity.

\subsection{The algebraic system}

Substitution of the representation \eq{Cpq} into the 16 relations \eq{eq:cond1}--\eq{eq:cond16} leads to a linear algebraic system with respect to the variables $\bC = \Big(C_{11}, C_{12}, C_{13}, C_{14}, \ldots, C_{41}, C_{42}, C_{43}, C_{44}\Big)$, as follows
\beq
\bmA(\omega, \bk) \bC^T =0,
\label{alg_syst}
\eeq
where $\bmA(\omega, \bk)$ is a $16 \times 16$ matrix valued function of the radian frequency $\Go$ and the Bloch vector $\bk=(k_x,k_y)$. The dispersion equation is then written as
\beq
\det \bmA(\omega, \bk) = 0.
\label{disp_eq}
\eeq
This equation corresponds to Floquet-Bloch waves, propagating through a lattice-like system, which in turn represents a lower-dimensional model, with three-dimensional thin ligaments connecting the junction region replaced by one-dimensional segments along which we solve a fourth-order ordinary differential equation and impose the appropriate junction conditions. 

\subsection{Dispersion surfaces and Dirac cones}

We consider the dispersion surfaces for the square lattice of elastic beams (both Euler-Bernoulli and Rayleigh beams) and identify frequencies corresponding to vertices of so-called ``Dirac cones'' (the vertex of the Dirac cone is also referred to as the Dirac point); in turn these correspond to intermediate frequency regimes, where dispersion surfaces become non-smooth and exhibit a degeneracy for structures possessing required symmetry. In particular, for elastic structured plates constrained by a doubly periodic array of rigid pins, the nature of Dirac cones and their parameters are discussed in the recent paper \cite{McPhedran_2015}.
The effects of rotational inertia, attributed to the Rayleigh beams, are given special attention.

\begin{figure}[!htb]
\centering
\includegraphics[width=60mm]{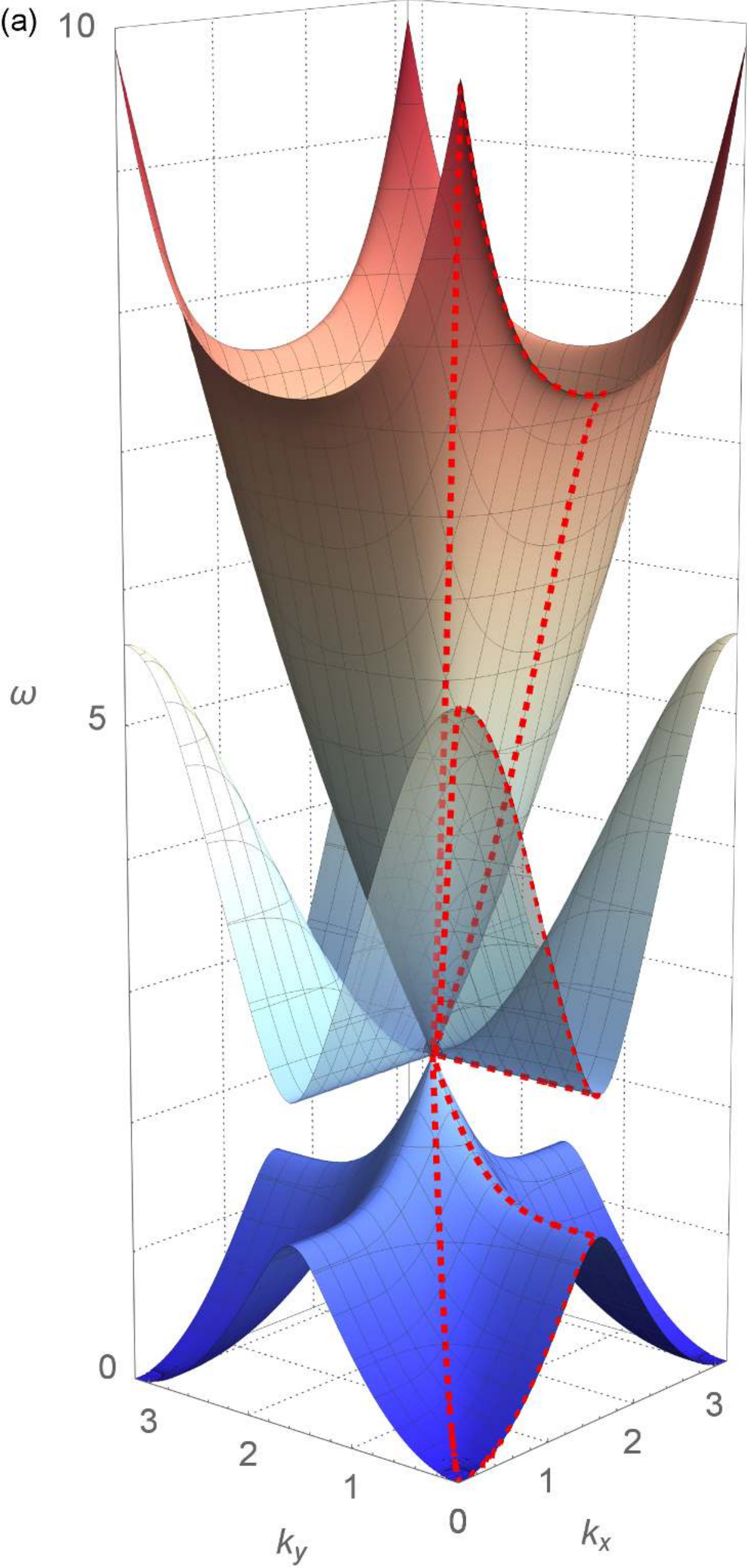}
\includegraphics[width=60mm]{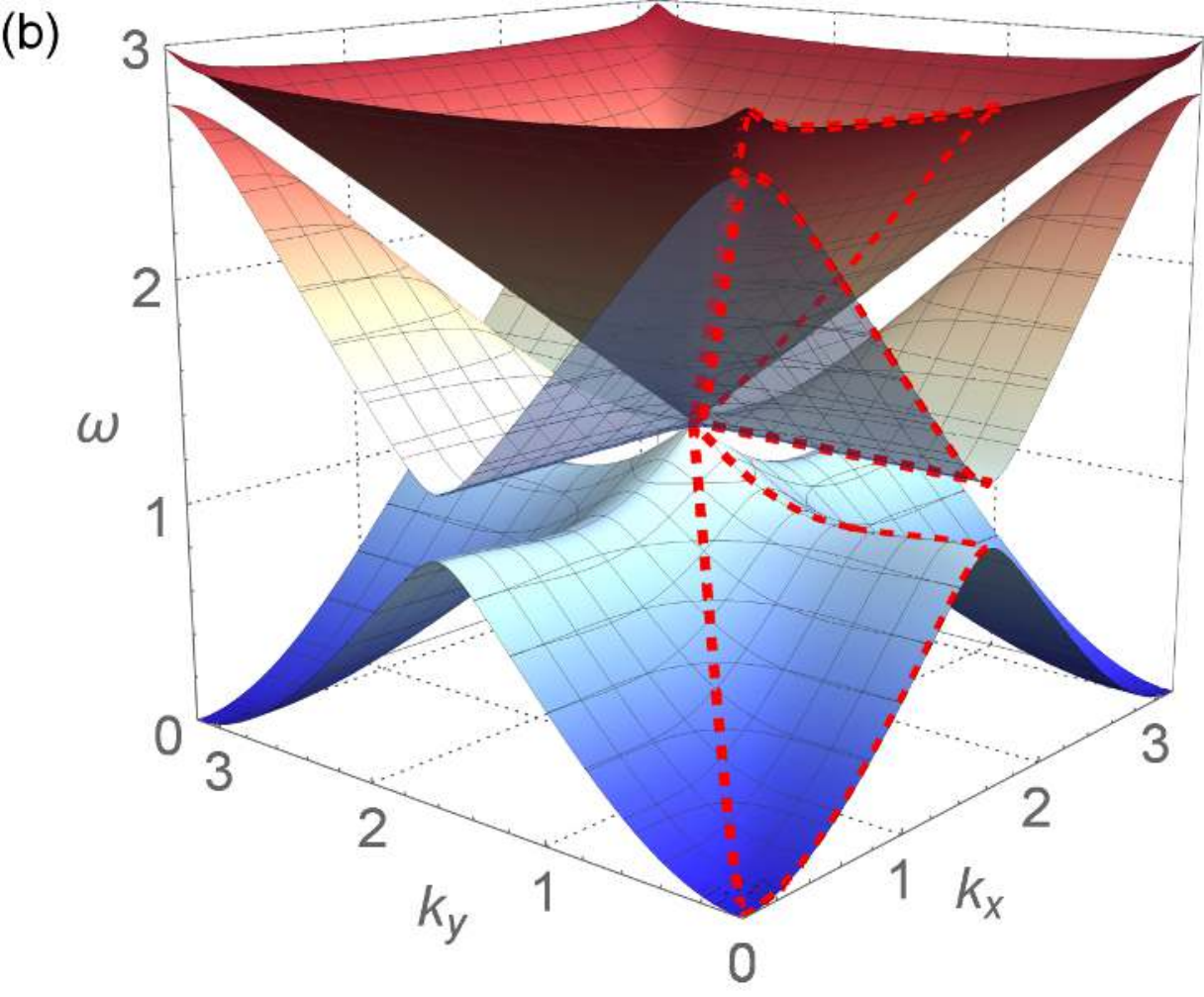}
\caption{\footnotesize Dispersion surface for the square Euler-Bernoulli beam structure (a), and for the square Rayleigh beam structure (b), in absence of pre-stress and elastic foundation. The chosen parameter values are $E=1$, $\rho=1$, $A=1$, $I=1$, $\ell_x=\ell_y=1$, $P_x=P_y=0$, $\beta=0$, $M_0=I_0=0$. We note that the effect of rotational inertia in (b) is significant: the first three dispersion surfaces shown here occur at much lower frequencies compared to the corresponding surfaces for the Euler-Bernoulli's beams as in part (a). Dirac cones are clearly visible in both cases. However, the dispersion profiles near the Dirac cone vertices are different for the Rayleigh beam structure and for the Euler-Bernoulli beam structure. The red dashed line highlights the cross-sectional diagram along the boundary of the irreducible Brillouin zone (compare with Fig.~\ref{figband2}).} 
\label{fig3d}
\end{figure}

In Fig.~\ref{fig3d} we give the solution of the dispersion equation (\ref{disp_eq}), represented in the form of the dispersion diagram, where the dispersion surfaces correspond to the roots $\Go = \Go(k_x, k_y)$. In the computations shown in this figure, it is assumed that the pre-stress and the Winkler foundation are absent. We point out at the multiple roots $\Go$, where a  dispersion surface becomes non-smooth, and appears as a cone in the space $(k_x, k_y, \omega)$.  Following \cite{McPhedran_2015} we refer to these conical surfaces as ``Dirac cones'', and the radian frequency $\Go$ corresponding to the vertex of the cone will be identified as a resonant frequency for a substructure of a certain type.
Fig.~\ref{fig3d}a shows the dispersion diagram for Floquet-Bloch waves in the lattice of Euler-Bernoulli beams. Fig.~\ref{fig3d}b presents the results for the network of the Rayleigh beams, possessing the rotational inertia. 

One important characteristic feature of these dispersion surfaces is the presence of multiple roots of the dispersion equation, represented as the vertex of the Dirac cone.
The Dirac point regime is essentially the dynamic regime and cannot be covered in the framework of the low-frequency homogenisation approximation. Also, such a dynamic regime has never been studied for multi-scale systems consisting of the Rayleigh beams, where the effects of rotational inertia bring new and exciting features in the dynamic response of the overall elastic system.

Fig.~\ref{fig3d} also shows the flat bands at relatively low frequencies, which correspond to zero group velocity waves, i.e.\ standing waves.

The dispersion equation characterising the Floquet-Bloch waves is a transcendental equation, and it is solved for non-negative $\omega$, when $(k_x, k_y)$ is in the Brillouin zone of the reciprocal space.
In particular, if the rotational inertia, the pre-stress and the stiffness of the elastic foundation are all replaced by zero, i.e. the Rayleigh beam becomes the classical Euler-Bernoulli beam, the dispersion equation \eq{disp_eq} is simplified to the form
\begin{multline}
\sin \left(2 \kappa \ell\right)
\left[\cos (2 k_x \ell) + \cos (2 k_y \ell) - 2 \cos \left(2 \kappa \ell\right)\right]
\left[\cos (2 k_x \ell) - \cosh \left(2 \kappa \ell\right)\right]
\left[\cos (2 k_y \ell) - \cosh \left(2 \kappa \ell\right)\right] \\
-\sinh \left(2 \kappa \ell\right)
\left[\cos (2 k_x \ell) + \cos (2 k_y \ell) - 2 \cosh \left(2 \kappa \ell\right)\right]
\left[\cos (2 k_x \ell) - \cos \left(2 \kappa \ell\right)\right]
\left[\cos (2 k_y \ell) - \cos \left(2 \kappa \ell\right)\right] = 0,
\end{multline}
where $\kappa = \sqrt{\omega} \sqrt[4]{\frac{\rho A}{EI}}$
and the corresponding dispersion diagram is shown in Fig.~\ref{fig3d}a.

{\bf Resonant modes.} For a simplified configuration where the pre-stress and the Winkler foundation are absent, the frequencies of Dirac cone vertices and the corresponding vibration modes are identified as the natural frequencies and the eigenmodes of a simply supported beam, respectively

\begin{equation}
\omega_n = \frac{n^2 \pi^2}{\ell^2} \sqrt{\frac{EI}{\rho A}} \qquad \text{for Euler-Bernoulli beam},
\end{equation}

\begin{equation}
\omega_n = \frac{n^2 \pi^2}{\ell} \sqrt{\frac{EI}{\rho (A\ell^2+n^2\pi^2I)}} \qquad \text{for Rayleigh beam}.
\end{equation}

In the general case, which includes pre-stress and non-zero stiffness of the Winkler foundation, the frequencies in question are given by:

\begin{equation}
\omega_n = \frac{1}{\ell} \sqrt{\frac{n^4\pi^4EI+n^2\pi^2P\ell^2+\beta\ell^4}{\rho (A\ell^2+\alpha n^2\pi^2I)}},
\label{simsupbeam}
\end{equation}
where $\alpha=1$ for Rayleigh beams and $\alpha=0$ for Euler-Bernoulli beams.

In the quasi-static regime, when the frequency $\Go$ is small, and for small values of $|\bk|$, the square lattice consisting of Euler-Bernoulli beams or Rayleigh beams gives a similar overall response as the result of the homogenisation procedure (see, for example, \cite{Panasenko2005}). 
For higher frequencies the dynamic response is anisotropic and frequency sensitive. Also the effect of the rotational inertia, which is present in the Rayleigh beam structure, becomes important, and consequently the dispersion surfaces representing structures consisting of the Rayleigh beams and structures consisting of the Euler-Bernoulli beams, become very different, as seen in Fig.~\ref{fig3d}.
First, Floquet-Bloch waves in a Rayleigh beam structure would show zero group velocity at much lower frequencies compared to the similar structure of the Euler-Bernoulli beams. Second, there are significant differences on the slowness contour diagrams, which will be discussed in the text below.

\begin{figure}[!htb]
\centering
\includegraphics[width=150mm]{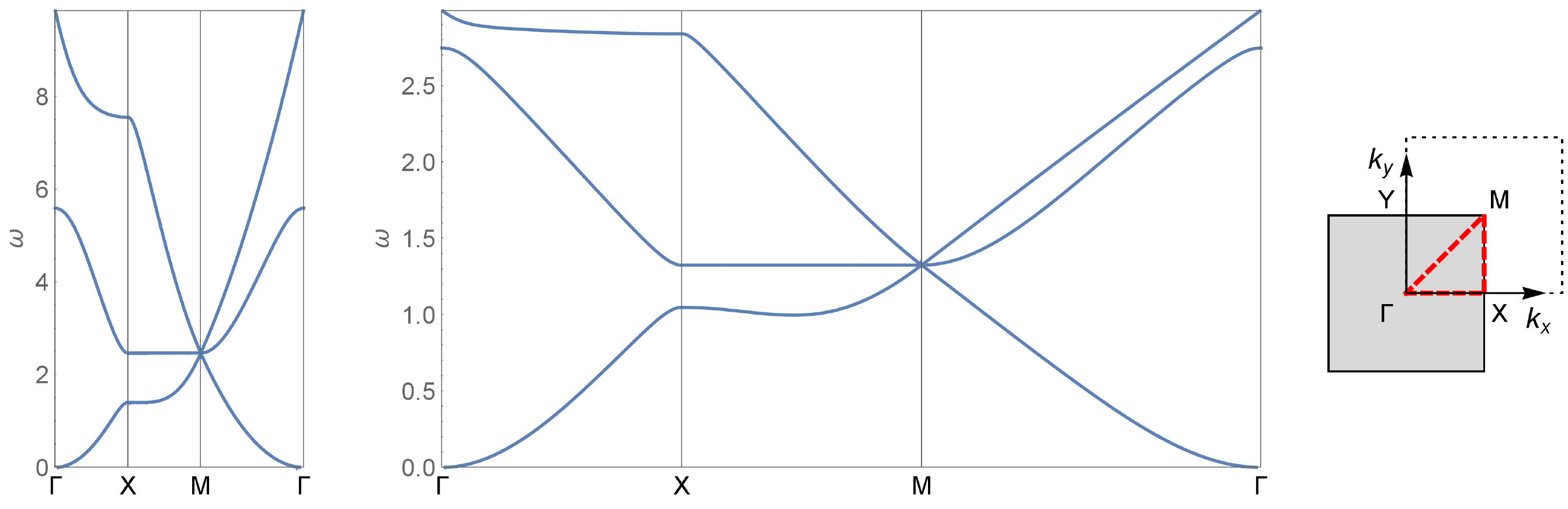}
\caption{\footnotesize The cross-sectional dispersion diagrams along the boundary of the irreducible Brillouin zone, for Floquet-Bloch waves in the networks of the Euler-Bernoulli beams (a) and the Rayleigh beams (b). The inset on the right shows the contour $\GG X M Y$ within the first Brillouin zone in the elementary cell of the reciprocal lattice, and the irreducible Brillouin zone (red dashed line); the dotted square corresponds to the computational window chosen to draw the dispersion surfaces in Fig.~\ref{fig3d}.}
\label{figband2}
\end{figure}

Fig.~\ref{figband2} complements the three-dimensional dispersion surfaces by the cross-sectional plots along the boundary $\GG X M Y$ of the first Brillouin zone in the elementary cell of the reciprocal lattice, shown in the inset of this figure.
The flat bands on these cross-sectional diagrams show the standing waves regimes, whereas the triple crossing points represent the vertices of the Dirac cones.

We refer to the notion of ``metamaterials'', used for multi-scale predesigned structures, which posses the required properties, in many cases unexpected and unusual. In particular, we discuss the connection between the rotational inertia in Rayleigh beams periodic networks and the effects of dynamic anisotropy and negative refraction.

\subsection{Slowing down the dynamic response of the beam lattice}

Our model demonstrates that the replacement of the Euler-Bernoulli beams by the Rayleigh beams in the square elastic lattice shifts the dispersion surfaces into the lower frequency range, as shown in Figs.~\ref{fig3d} and \ref{figband2}. Physically it implies that the dynamic response of the Rayleigh beam lattices is ``slower'' than the dynamics response of a similar lattice, but consisting of the Euler-Bernoulli beams. This feature applies to frequencies of standing waves as well as frequencies corresponding to the Dirac points.
It is also noted that the above mentioned phenomenon is typical for the higher frequency range, but in the quasi-static regime the Eurler-Bernoulli and the Rayleigh beam structures become indistinguishable in the limit as $\Go \to 0$.

\section{Slowness contours, dynamic anisotropy, standing waves}
\label{sec04}

Fig.~\ref{fig3d}a gives a set of three dispersion surfaces for waves in the Euler-Bernoulli beam structure, and its analogue for the Rayleigh beams is given in Fig.~\ref{fig3d}b. The difference is striking, especially for higher frequencies regime.
In Figs.~\ref{slowness_contours}--\ref{slowness_contours3} we consider these surfaces individually, together with the corresponding slowness contours, i.e. the cross-sections of the three-dimensional surface at fixed frequency.

The slowness contours (also referred to as isofrequency contours) are used to describe the dynamic anisotropy of the structured medium. A nice exposition of this approach for vector problems of elasticity in periodic lattices is presented in the paper \cite{Fleck2006}.
The case we study involves a scalar fourth-order problem, and we focus on analysis of degeneracies and irregularities of dispersion surfaces (in particular, of Dirac cones) in conjunction with the slowness contours.

\subsection{Dynamic anisotropy and negative refraction: Rayleigh beams versus Euler-Bernoulli beams at low and intermediate frequencies}

\begin{figure}[!htb]
\centering
\includegraphics[width=140mm]{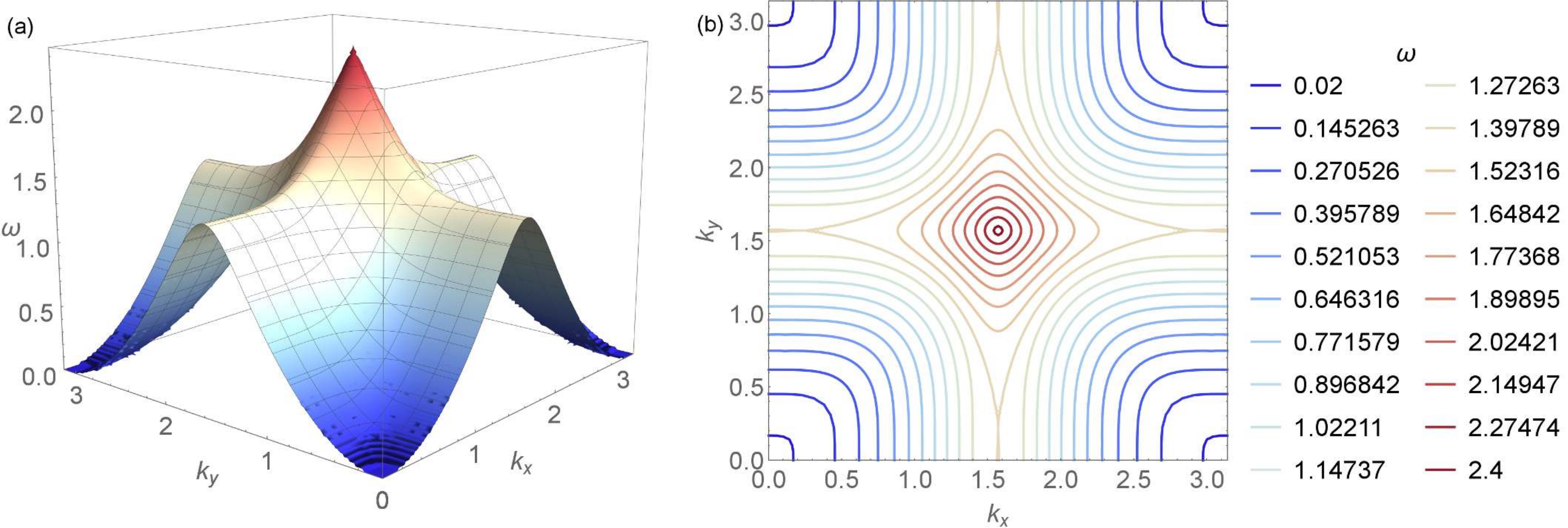} \\[6mm]
\includegraphics[width=140mm]{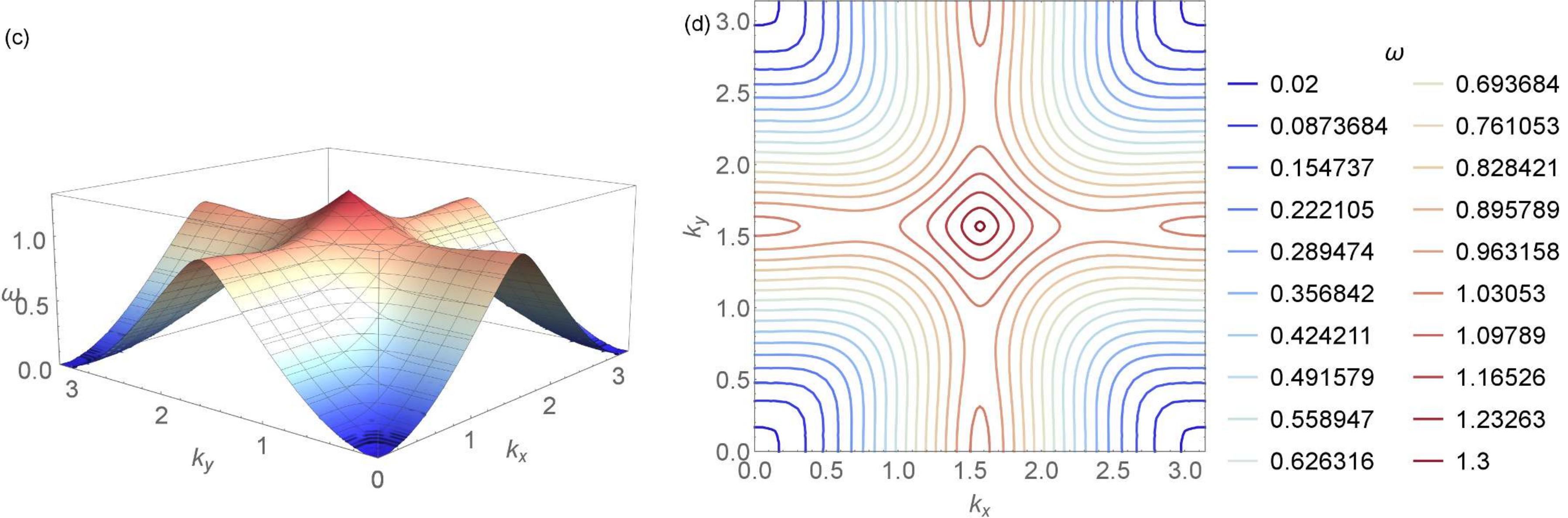}
\caption{\footnotesize
First dispersion surface and the corresponding isofrequencies contours for the Euler-Bernoulli beam square lattice (parts (a) and (b)) and for the square lattice of the Rayleigh beams (parts (c) and (d)). The Dirac cone is shown in both configurations. The slowness contours around the origin bound non-convex domains in the diagram (d), for the Rayleigh beams, in contrast with the diagram (b), corresponding to the Euler-Bernoulli beams. }
\label{slowness_contours}
\end{figure}

The lowest dispersion surfaces (corresponding to so-called acoustic bands), and their slowness contour are presented in Fig.~\ref{slowness_contours}. 
In terms of effective dynamic response for low frequency, the square lattice of flexural beams (both the Euler-Bernoulli and the Rayleigh) gives an anisotropic response in contrast to the membrane problem, where the square lattice of harmonic springs would respond isotropically for the low frequencies, and hence the dispersion surface for such a structured membrane would be a cone with the circular cross-section.
This difference between flexural plates and elastic membranes is expected. However, there are additional features of flexural Floquet-Bloch waves, which are rather counter-intuitive.

As shown in Fig.~\ref{slowness_contours}, the slowness contours have a shape (squares with rounded corners) which indicates preferential directions along the coordinate axes, as well as oblique directions, both for the Rayleigh and the Euler-Bernoulli beams, used as constituents of the lattice.
As expected, for low frequencies the dispersion surfaces on parts (a) and (c) look similar, and the group velocity at the zero frequency is equal to zero. It is also not surprising that the slowness contours indicate a directional preference in the low frequencies interval, aligned with the principal axes of the lattice, i.e. the waves will propagate {\it along the beams}.

The interesting feature is observed for higher frequencies, which are close to the Dirac cone: the directional preferences change and the waves will propagate at an angle to the principal directions of the original lattice. 

As demonstrated in the earlier paper \cite{PM_2014}, the main difference between the periodic Euler-Bernoulli beam and the Rayleigh beam occurs at higher frequencies. This difference in a two-dimensional periodic configuration is clearly shown in   Fig.~\ref{slowness_contours}, in the intermediate regime, close to standing waves (which possess zero group velocity), where the slowness contours for the Euler-Bernoulli beam and the Rayleigh beam are convex (Fig.~\ref{slowness_contours}b) and non-convex (Fig.~\ref{slowness_contours}d), respectively.


The latter observation has a special significance in applications linked to transmission problems and structured interfaces, which are ``built'' of the square lattice of beam-like elastic ligaments. Namely, the non-convex slowness contours for the lattice consisting of Rayleigh beams imply the effect of the negative refraction. Hence, compared to the classical Euler-Bernoulli beams, the square lattice of the Rayleigh beams gives both a strong anisotropy and a negative refraction.

\subsection{Dispersion and standing waves:  Rayleigh beams versus Euler-Bernoulli beams at higher frequencies}

\begin{figure}[!htb]
\centering
\includegraphics[width=140mm]{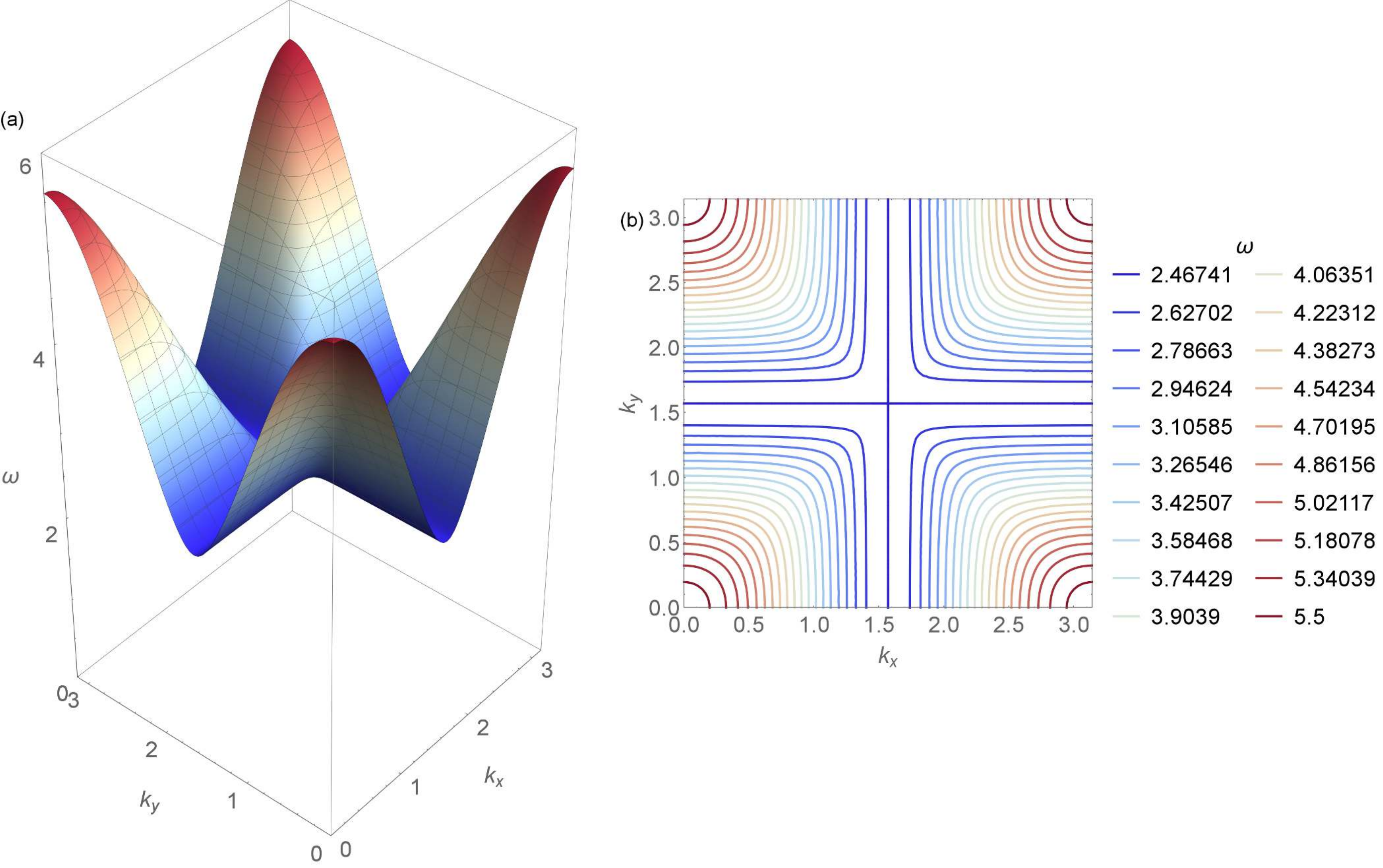} \\[6mm]
\includegraphics[width=140mm]{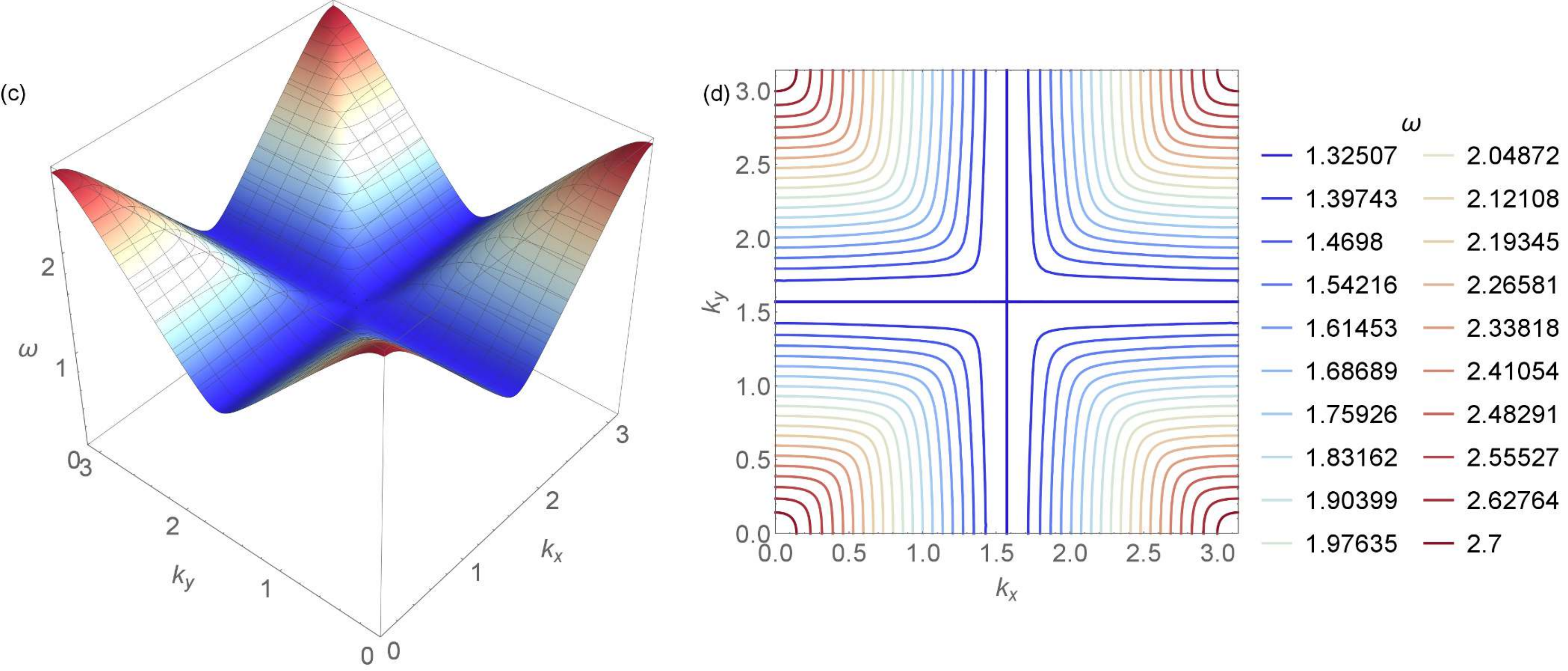}
\caption{\footnotesize
The second dispersion surface, including the Dirac cone, presented for the Floquet-Bloch waves in the case of the Euler-Bernoulli beam square lattice (parts (a) and (b)) and of the Rayleigh beam square lattice (parts (c) and (d)). Preferential directions along the coordinate axes are clearly identified.}
\label{slowness_contours2}
\end{figure}

The second set of dispersion surfaces shown in Fig.~\ref{slowness_contours2}, both for the lattice of the Euler-Bernoulli beams and the lattice of the Rayleigh beams, shows the existence of standing waves along the beams, at the Dirac point frequency, corresponding to the troughs shown in the dispersion surface. 
Also the preferential directions are, as expected, along the beams comprising the lattice.

\begin{figure}[!htb]
\centering
\includegraphics[width=140mm]{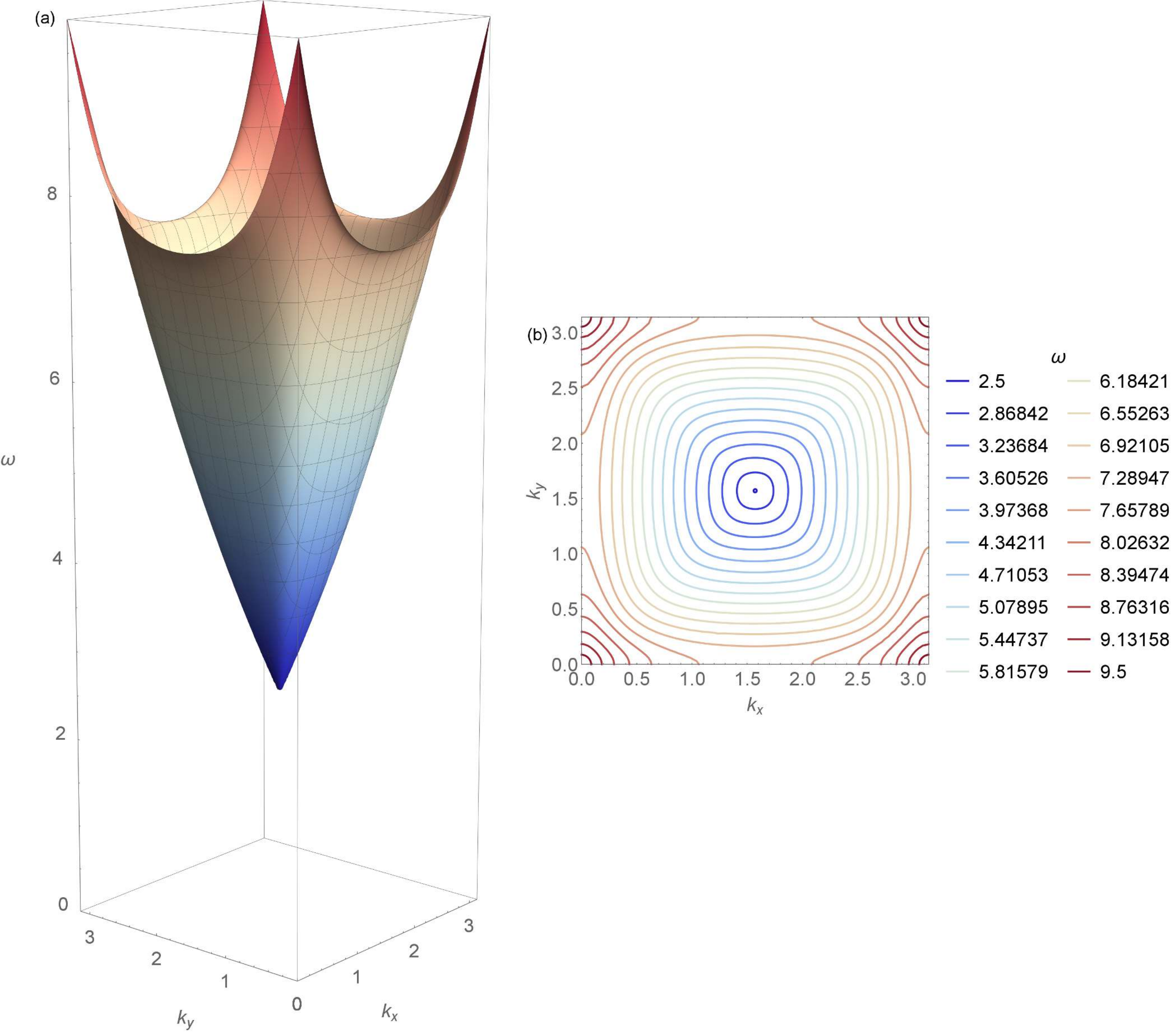} \\[6mm]
\includegraphics[width=140mm]{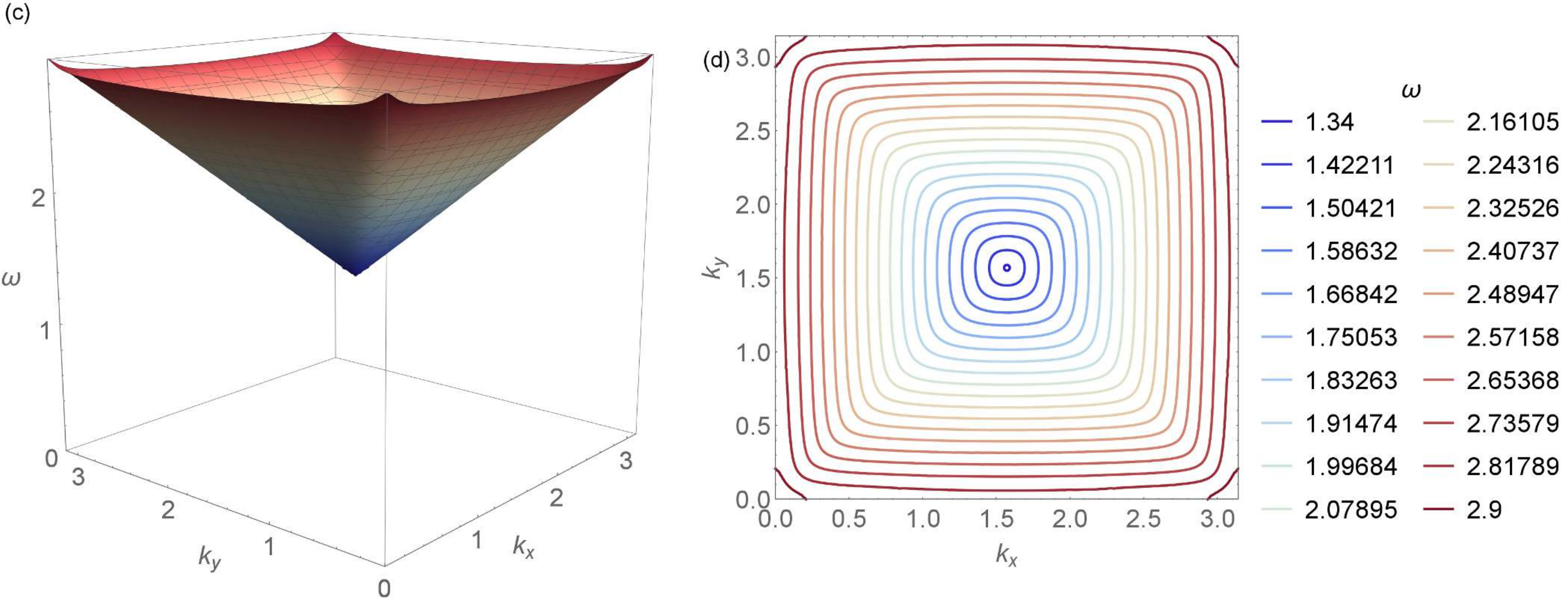}
\caption{\footnotesize
The third dispersion surface representing  a part of the nested Dirac cone structure for the case of the square network of the Euler-Bernoulli beams (parts (a) and (b)) as well as the Rayleigh beams (parts (c) and (d)). As expected, significant differences in dispersion of Floquet-Bloch waves are observed for the Rayleigh and Euler-Bernoulli beams at higher frequencies. }
\label{slowness_contours3}
\end{figure}

Finally, the third set of upper dispersion surfaces, representing the Dirac cones, is shown in Fig.~\ref{slowness_contours3}. These cones have different opening angles, with the cone in Fig.~\ref{slowness_contours3}a, which corresponds to Euler-Bernoulli beams, being significantly sharper than the cone in Fig.~\ref{slowness_contours3}c, which corresponds to Rayleigh beams. This observation is in line with the property of the group velocity of a flexural wave in the Euler-Bernoulli beam to grow to infinity with the increase of the frequency. On the contrary, the group velocity for flexural waves in the Rayleigh beam is bounded.

\clearpage

\subsection{Standing waves}

By definition, standing waves are those waves whose group velocity is equal to zero, i.e. such waves do not carry the bulk energy, and on the dispersion diagrams shown in Figs.~\ref{fig3d} and
\ref{figband2}, $\nabla \Go(k_x, k_y) =0$ at the points corresponding to such standing waves.

We pay special attention to those points on the dispersion diagram, which are on the segment $X M$ at frequencies corresponding to flat bands in  Fig.~\ref{figband2}. The corresponding normalised frequency for the Euler-Bernoulli beam lattice is $\Go_E \simeq 2.46$ and for the Rayleigh beam lattice it is $\Go_R \simeq 1.32$.
Those values are approximated analytically by the formula (\ref{simsupbeam}), where $n=1$, and the parameter $\alpha$ is chosen as
$\alpha =0$ for Euler-Bernoulli beams and $\alpha =1$ for Rayleigh beams.

\begin{figure}[!htb]
\centering
\includegraphics[width=135mm]{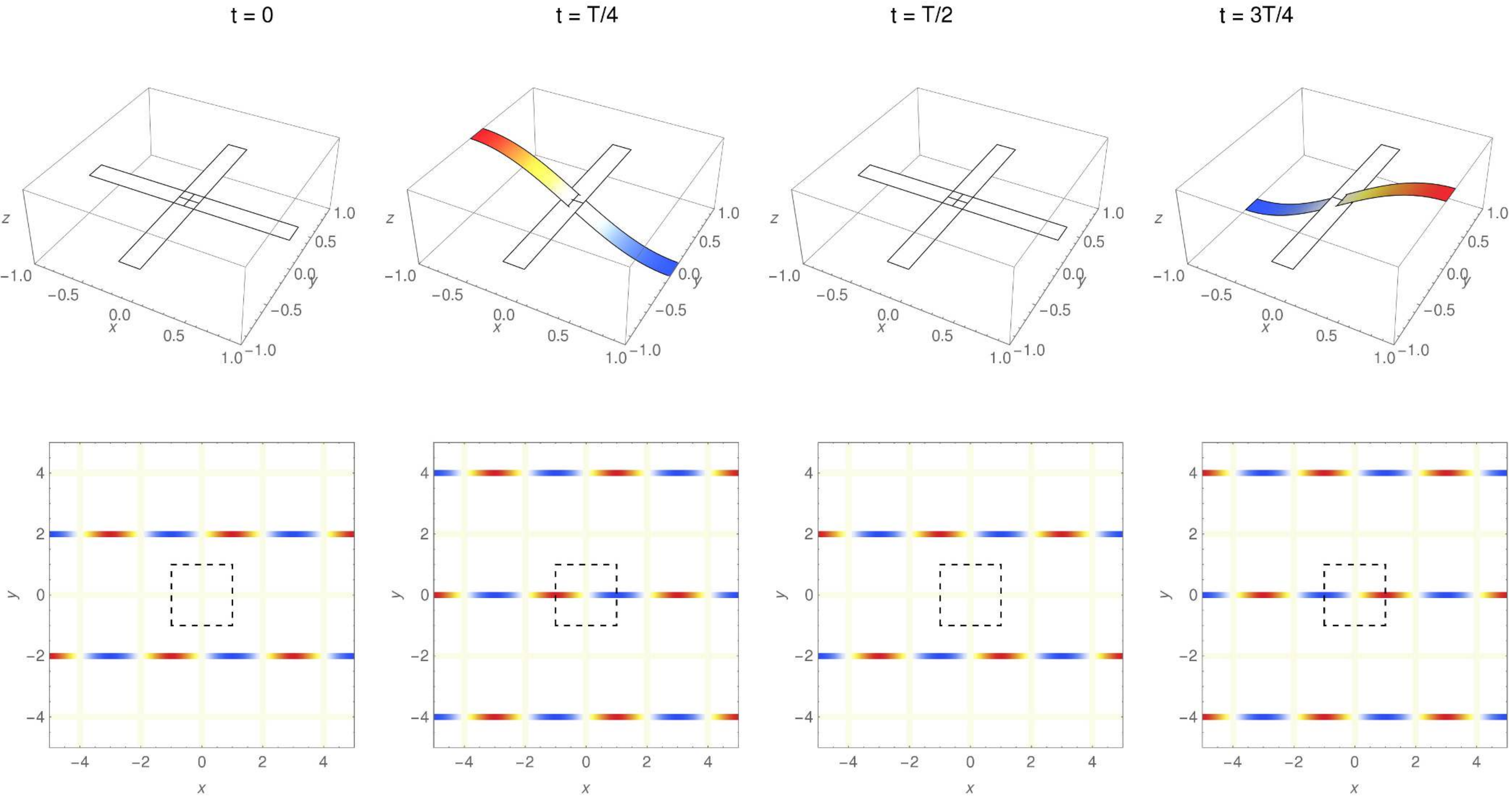}
\caption{\footnotesize
Standing wave in the network of Rayleigh beams for the angular frequency $\omega_R \simeq 1.32$ and wave vector $k_x=\pi/2$, $k_y=\pi/4$. Four different configurations are shown, at times $t=\{0,T/4,T/2,3T/4\}$, where $T$ is the period. In each elementary cell, containing a vibrating beam, the vibration mode resembles a simply supported flexural beam. Similar vibration modes occur for beams oriented in the vertical direction.}
\label{figmodst1RA}
\end{figure}

\begin{figure}[!htb]
\centering
\includegraphics[width=135mm]{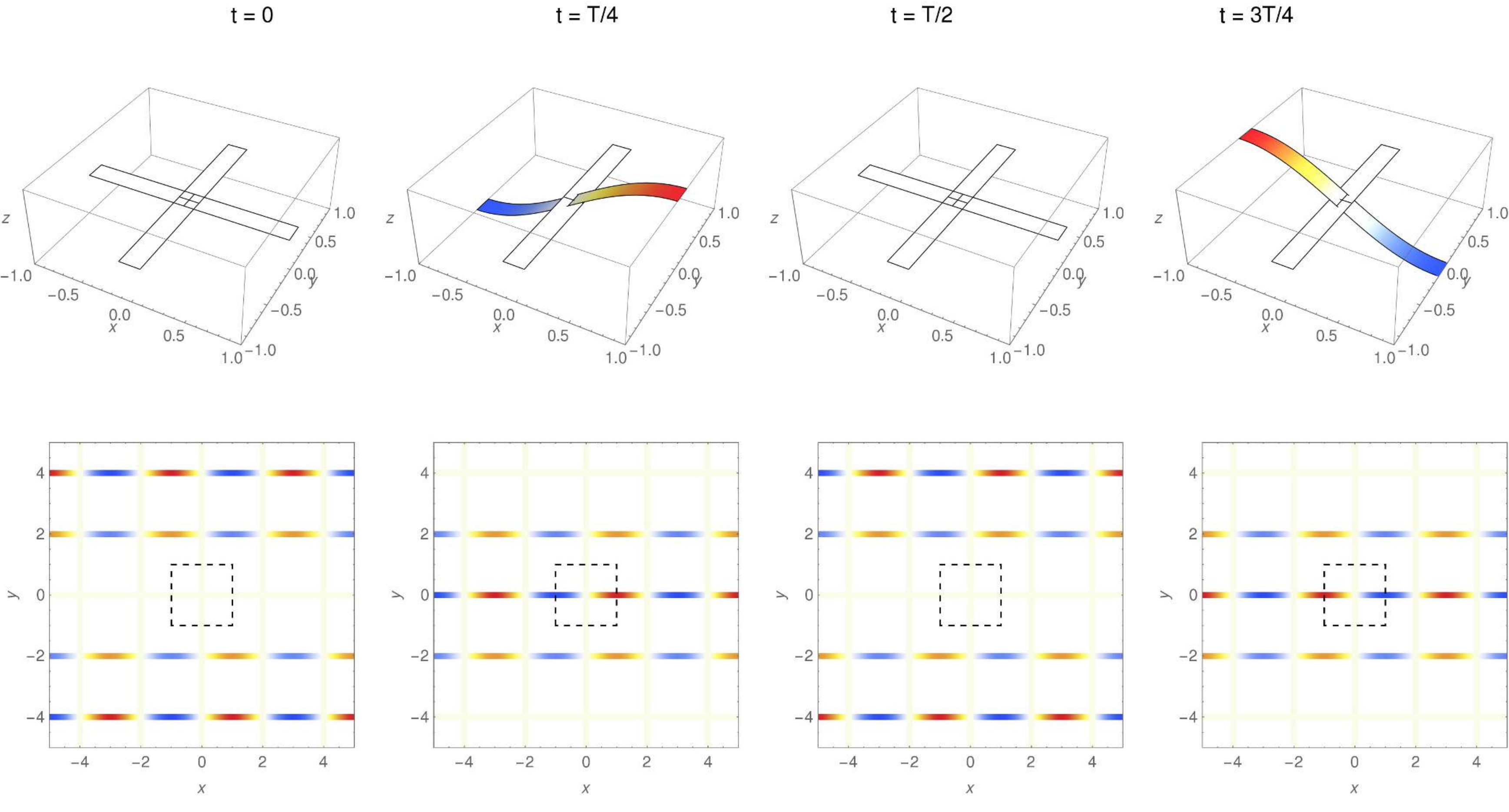}
\caption{\footnotesize
Modulation of standing wave in the network of Rayleigh beams for the angular frequency $\omega_R \simeq 1.32$ and wave vector $k_x=\pi/2$, $k_y=\pi/8$. Four different configurations are shown, at times $t=\{0,T/4,T/2,3T/4\}$, where $T$ is the period. In each elementary cell, containing a vibrating beam, the vibration mode resembles a simply supported flexural beam. Compared to Fig.~\ref{figmodst1RA}, the phase shift in the vertical direction has changed, which has resulted in a different modulation of standing flexural waves.}
\label{figmodst2RA}
\end{figure}

\begin{figure}[!htb]
\centering
\includegraphics[width=135mm]{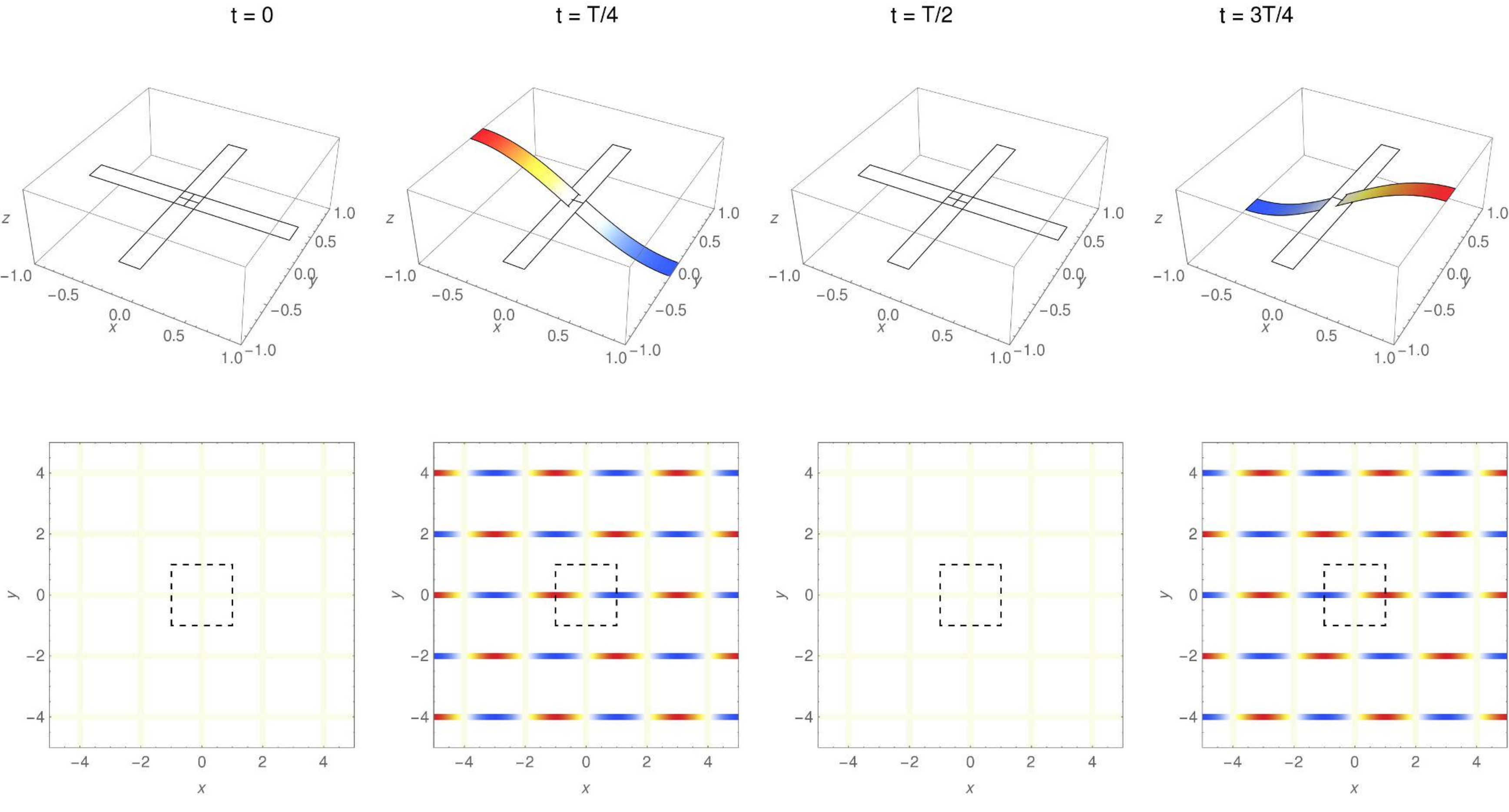} \\
\caption{\footnotesize The first eigenmode corresponding to the Dirac cone vertex, at the normalised angular frequency $\omega_R \simeq1.32$ and wave vector $k_x=\pi/2$, $k_y=\pi/2$. Four different configurations are shown, at times $t=\{0,T/4,T/2,3T/4\}$, where $T$ is the period. This waveform shows waves aligned with the $x-$axis; similar waveform (but rotated through the angle of $\pi/2$) corresponds to the second eigenmode with the waves aligned along the $y-$axis. }
\label{figmodst3RA}
\end{figure}

\begin{figure}[!htb]
\centering
\includegraphics[width=135mm]{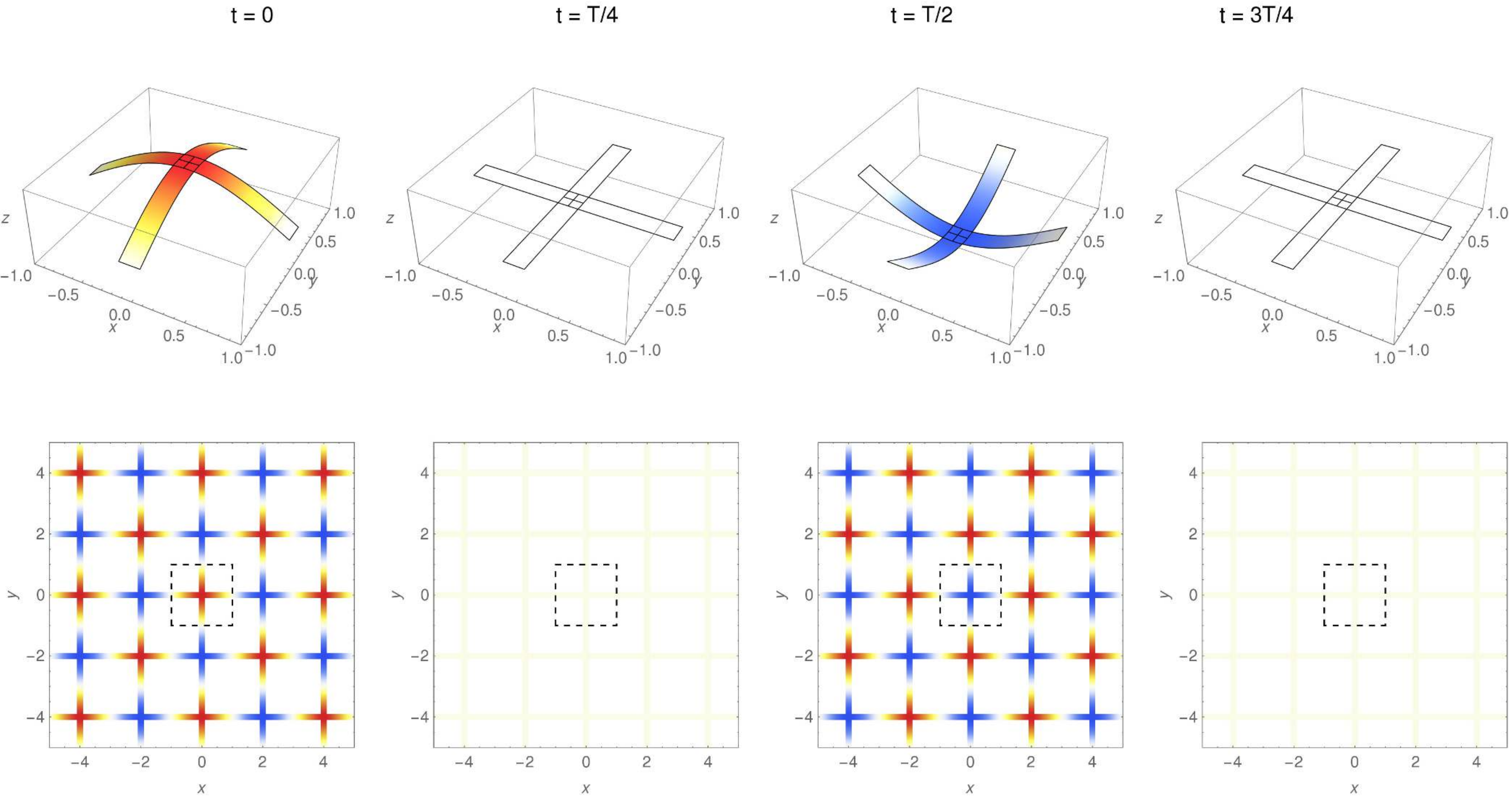}
\caption{\footnotesize The third eigenmode corresponding to the Dirac cone vertex, at the normalised angular frequency $\omega_R \simeq1.32$ and wave vector $k_x=\pi/2$, $k_y=\pi/2$. Four different configurations are shown, at times $t=\{0,T/4,T/2,3T/4\}$, where $T$ is the period. This waveform shows the directional preferences along the coordinated axes, with $x-$ and $y-$ directions being equivalent. Global modulation, with axes inclined at $\pi/4$ to the $x-$ and $y-$ axes is also observed.}
\label{figmodst4RA}
\end{figure}

In Figs.~\ref{figmodst1RA}--\ref{figmodst4RA} we give examples of standing waves at frequencies mentioned above for the Rayleigh beam square lattice. Whenever we show the modes, corresponding to the vibration of the horizontal beams, the same type of standing waves exists for beams vibrating in the vertical direction, at the same frequency. This follows from the equivalence of the $x$ and $y$ directions for a square lattice of beams.
We also note that similar shapes of standing waves occur for the Euler-Bernoulli beams, but at a higher frequency, as discussed in the previous section.

The frequency $\Go_R$ is maintained for the standing waves along the dispersion path $XM$, but the vibration modes changes when $k_y$ changes while $k_x=\pi/2$ remains fixed. For the cases when $k_y=\pi/4, \pi/8$, these vibration modes are given in Figs.~\ref{figmodst1RA} and \ref{figmodst2RA}, respectively.
Although the vibration modes of horizontal beams shown in Figs.~\ref{figmodst1RA} and \ref{figmodst2RA} appear to be identical, there is a difference in the phase shift along the vertical axis for these two types of standing waves.

Figs.~\ref{figmodst3RA} and \ref{figmodst4RA} illustrate the waveforms corresponding to the Dirac cone vertex, at the normalised angular frequency $\omega_R \simeq 1.32$ and wave vector $k_x=\pi/2$, $k_y=\pi/2$. These waveforms represent three eigenmodes, two of which are the waves along the coordinate axes (as shown in Fig.~\ref{figmodst3RA}) and the mirror symmetric standing wave displayed in Fig.~\ref{figmodst4RA}.

\clearpage

\section{The rectangular lattice}

\begin{figure}[!htb]
\centering
\includegraphics[width=60mm]{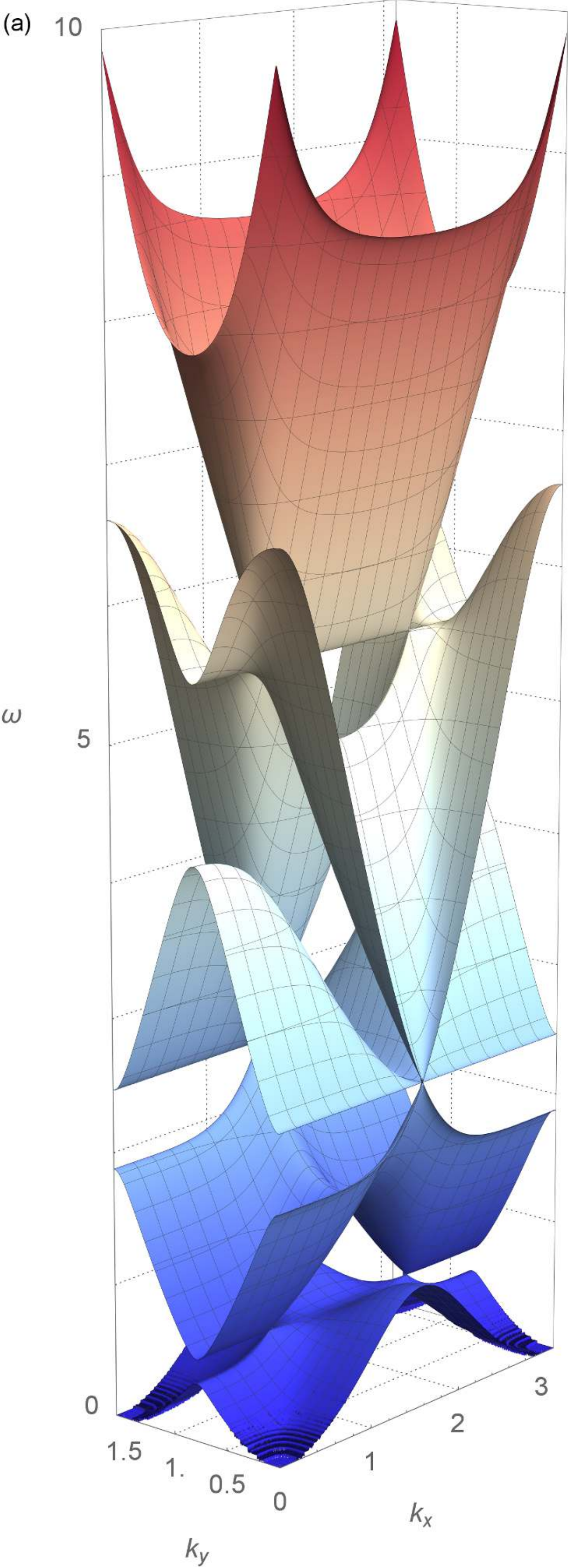}
\includegraphics[width=60mm]{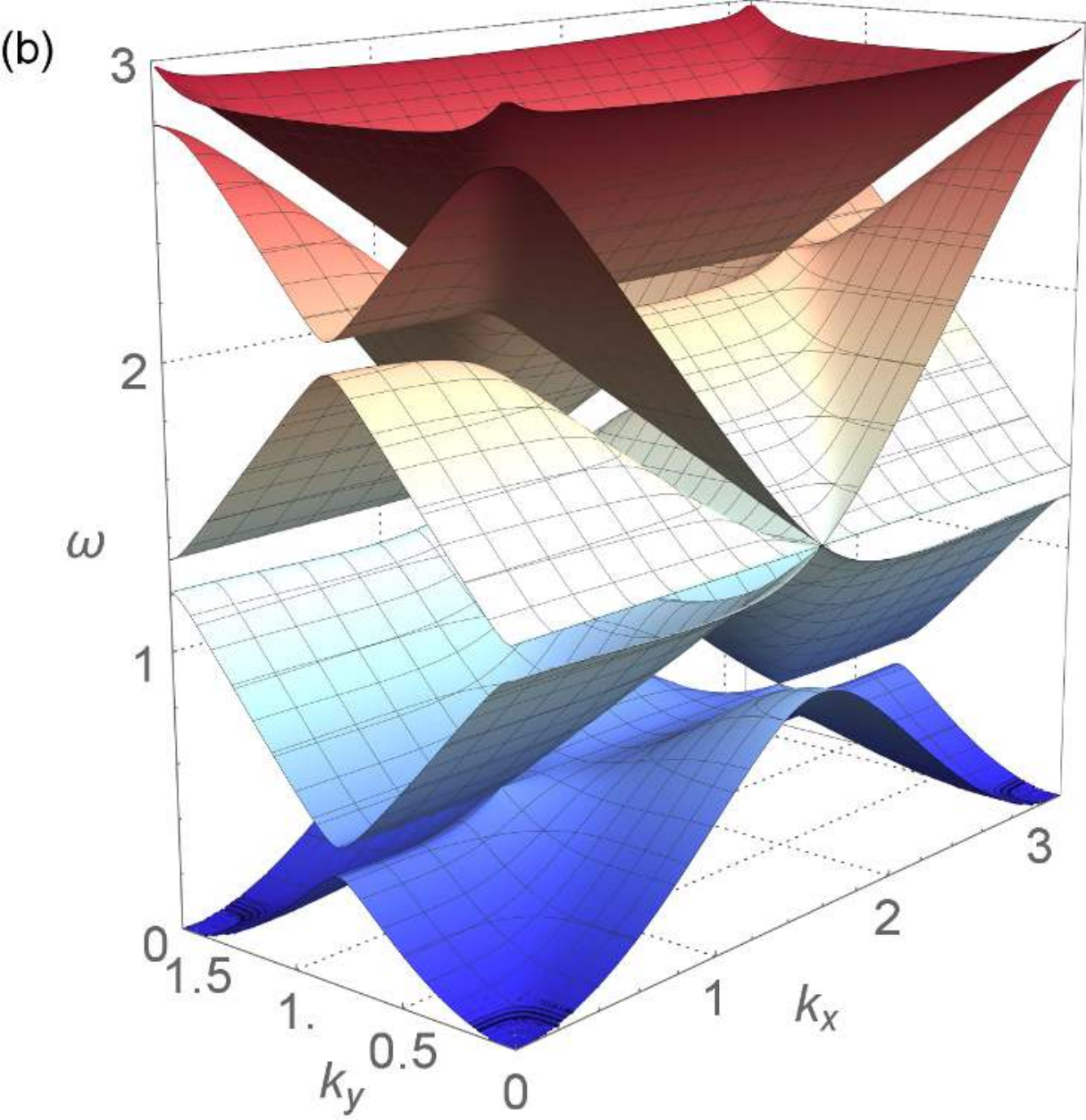}
\caption{\footnotesize Dispersion surface for the rectangular Euler-Bernoulli beam structure (a), and for the rectangular Rayleigh beam structure (b). We note that the effect of rotational inertia in (b) is significant: the dispersion surfaces shown here occur at much lower frequencies compared to the corresponding surfaces for the Euler-Bernoulli's beams as in part (a). The Dirac cone vertex, shown in Fig.~\ref{fig3d} for the square lattice, has been split into two Dirac cones, with their vertices being shifted to the boundary of the Brillouin zone. We also note the presence of a ``Dirac edge'' that connects two double roots of the dispersion equation at a lower frequency, which correspond to directionally localised standing waves.}
\label{figrect3d}
\end{figure}

If the square lattice is replaced by a rectangular network of beams with the integer ratio of the characteristic length in the elementary cell, we lose some of the Dirac cones on the dispersion surfaces constructed for the elastic Floquet-Bloch waves. This is illustrated in Fig.~\ref{figrect3d}, which presents the dispersion diagrams for rectangular networks of the Euler-Bernoulli (part (a)) and of the Rayleigh (part (b)) beams, for the aspect ratio of the characteristic length in the elementary cell $d_y/d_x=2$. The Dirac cone vertex, shown in Fig.~\ref{fig3d}, has been split into two Dirac cones, with their vertices being shifted to the boundary of the Brillouin zone. We also note the presence of a ``Dirac edge'' that connects two double roots of the dispersion equation at a lower frequency, as displayed in Fig.~\ref{figrect3d}, which correspond to directionally localised standing waves.
As expected, the network of the Rayleigh beams with rotational inertia lowers the typical frequencies  of propagating flexural waves.

\begin{figure}[!htb]
\centering
\includegraphics[width=140mm]{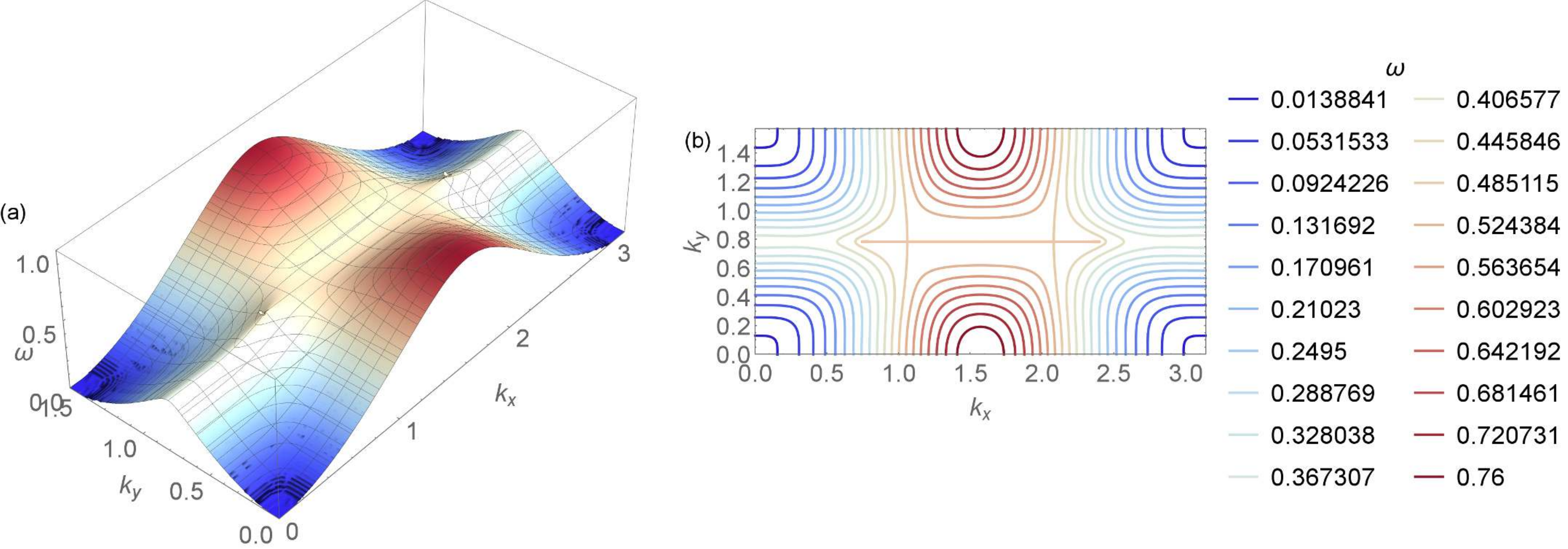}
\caption{\footnotesize The first dispersion surface (part (a)) and the corresponding slowness contours (part (b))
for the Floquet-Bloch waves in the rectangular network of the Rayleigh beams. Three types of stationary points are identified: points of local maxima (elliptic regime) and ``ridge'' points, which correspond to the hyperbolic and parabolic regimes, and hence directional localisation of the Floquet-Bloch waves. The first dispersion surface of the rectangular network of the Rayleigh beams has the lower frequency range than the one for the Euler-Bernoulli beams, but the structure of slowness contours is similar for both networks.
}
\label{figRECTLFRA}
\end{figure}

\begin{figure}[!htb]
\centering
\includegraphics[width=140mm]{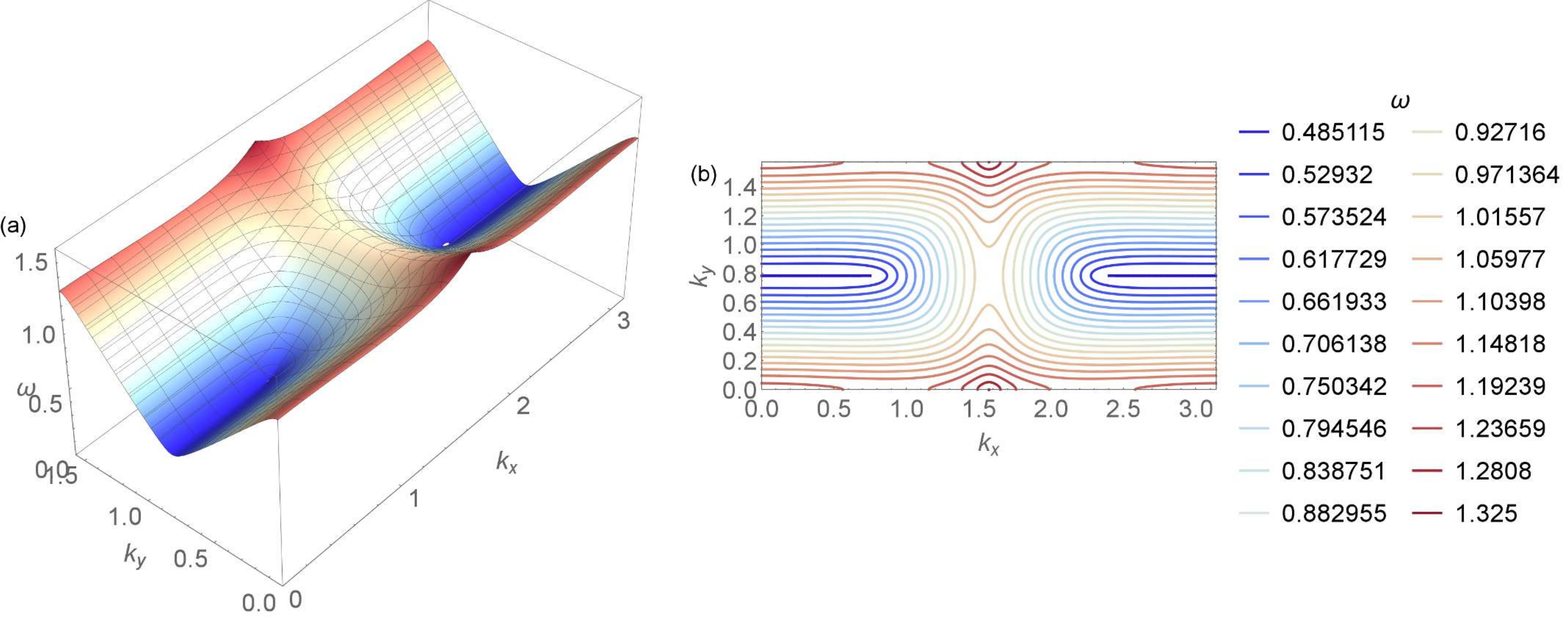}
\caption{\footnotesize The second dispersion surface (part (a)) and the corresponding slowness contours (part (b))
for the Floquet-Bloch waves in the rectangular network of the Rayleigh beams. The ``ridge'' type  stationary points are identified, which correspond to the hyperbolic and parabolic regimes, and hence directional localisation of the Floquet-Bloch waves. The second dispersion surface of the rectangular network of the Rayleigh beams has the lower frequency range than the one for the Euler-Bernoulli beams, and the slowness contours exhibit non-convexity compared to the slowness contour which represent the second dispersion surface for the Euler-Bernoulli network of beams.
}
\label{figRECTMFRA}
\end{figure}

\begin{figure}[!htb]
\centering
\includegraphics[width=140mm]{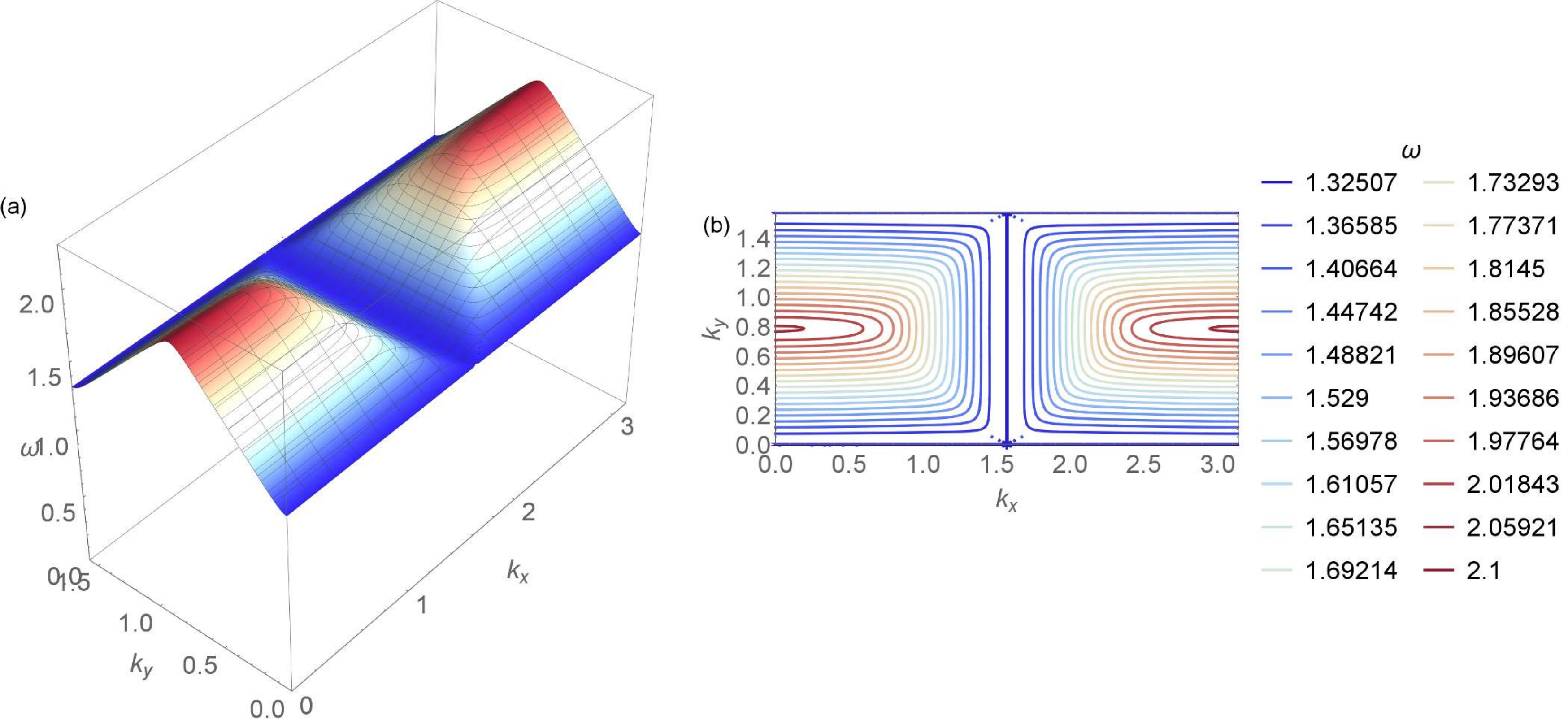}
\caption{\footnotesize The third dispersion surface (part (a)) and the corresponding slowness contours (part (b))
for the Floquet-Bloch waves in the rectangular network of the Rayleigh beams. The local maxima stationary points (elliptic regime) and the ``ridge'' type  stationary points are identified, in particular the ``ridge'' stationary points correspond to a parabolic regime, and hence a uni-directional localisation of the Floquet-Bloch waves. The third dispersion surface of the rectangular network of the Rayleigh beams has the lower frequency range than the one for the Euler-Bernoulli beams, but the slowness contours are similar to the slowness contour which represent the third dispersion surface for the Euler-Bernoulli network of beams.
}
\label{figRECTHFRA}
\end{figure}

Special attention is given to the first three dispersion surfaces, for which we also present the slowness contour diagrams (isofrequency maps), as shown in  the next three Figures \ref{figRECTLFRA}, \ref{figRECTMFRA}, \ref{figRECTHFRA}.
The first dispersion surface from Fig.~\ref{figrect3d}b and its slowness contour diagram are shown in Fig.~\ref{figRECTLFRA}.
A special feature here is the presence of saddle points as well as ``ridges'' on the dispersion diagram which indicate strong directional anisotropy for low frequency waves.
The second dispersion surface and the corresponding set of slowness contours are shown in Fig.~\ref{figRECTMFRA}, which exhibits several regimes associated with directional localisation of the Floquet-Bloch waves. In particular, these are linked to the ``ridge'' stationary points clearly visibly on the dispersion surface, and we especially mention the parabolic regimes, which correspond to a uni-directional localisation of the Floquet-Bloch waves.
The third dispersion surface and the corresponding slowness contours are shown in Fig.~\ref{figRECTHFRA}, which displays the local maxima stationary points, that correspond to standing waves in the elliptic regime. In addition we also observe the local minima region, which can be locally approximated by a parabolic cylinder, and in such regimes it is expected that the Floquet-Bloch waves will exhibit a uni-directional localisation.


\clearpage

\section{Forced network of Rayleigh beams}

\begin{figure}[!htb]
\centering
\includegraphics[width=100mm]{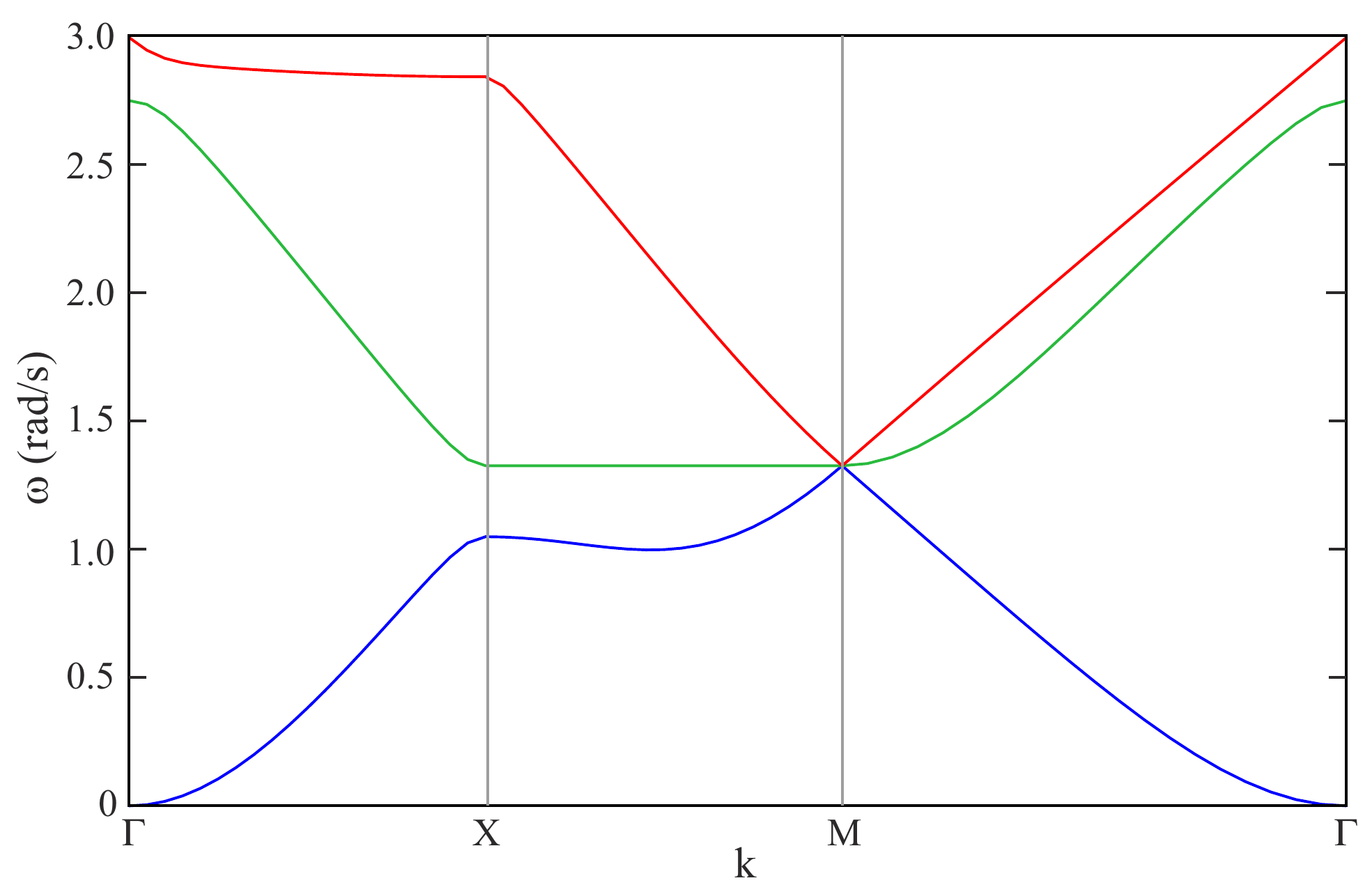}
\caption{\footnotesize The comparative finite element computation of the dispersion curves around the irreducible Brillouin zone for flexural Floquet-Bloch waves in the square network of the Rayleigh beams. The results of the finite element computation are identical to the analytical computation presented in Fig.~\ref{figband2}.}.
\label{figmodst4RA1}
\end{figure}

The analysis of the above sections, which has addressed dispersion properties of Floquet-Bloch waves in the periodic Rayleigh beam systems and, in particular, strong  dynamic anisotropy, will be used here to demonstrate applications to problems of forced lattice systems consisting of networks of Rayleigh beams.
The physical parameters are chosen to be identical to those, which were used in the earlier Section \ref{sec03}.
For comparison, we have produced the dispersion diagram in COMSOL Multiphysics with Floquet-Bloch conditions imposed on the boundary of the elementary square cell (see Fig.~\ref{figmodst4RA1}). As expected, the results of the finite element computation appear to be identical to the analytical computations presented in Fig.~\ref{figband2}b.

The new computations for a lattice loaded by a point force, perpendicular to the $(x, y)-$plane and applied to a junction of the square network of the Rayleigh beams, are produced in COMSOL in the frequency response mode.
We would like to add that the computations were not based on a standard, pre-programmed COMSOL package, since the Rayleigh beams are not covered by the standard COMSOL distribution. The equations of the Rayleigh beam were programmed in the form of two coupled ODEs as follows
\begin{align}
& M = -EI(u'' + \frac{\rho}{E}\omega^2 u), \\
& M'' +\rho A \omega^2 u = 0.
\end{align}

Moreover, in order to simulate an infinite lattice with a finite-size computational window and avoid wave reflection at the boundaries, Rayleigh and Euler-Bernoulli beams with damping were also programmed in COMSOL. This was obtained by replacing the Young modulus $E$ by a complex value, $E(1+i\eta)$. These beams were used to build a damping layer around the perimeter of the finite-size lattice, and the viscous parameter $\eta$ was chosen so that to minimize the wave reflection.

The results illustrate the predicted dynamic anisotropy and are discussed below.

\subsection{Uni-directional localisation}

\begin{figure}[!htb]
\centering
\includegraphics[width=90mm]{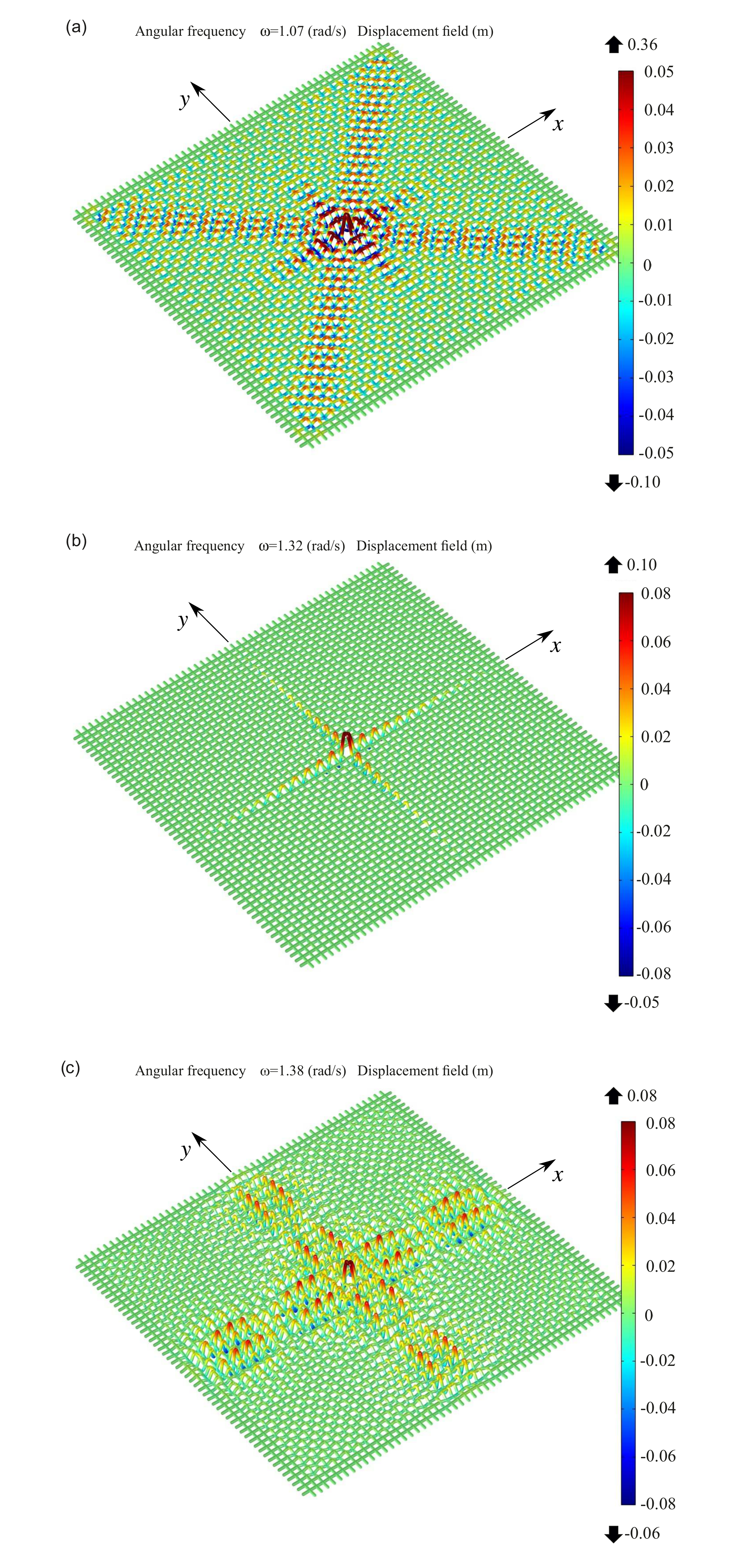}
\caption{\footnotesize
(a) The time-harmonic point-force load is applied to the network of Rayleigh beams at the normalised angular frequency of $1.07$. This corresponds to a sub-Dirac cone regime, where the predicted preferential directions are at $45^\circ$ with respect to the $x-$ and $y-$ axes and evanescent waveforms prevail. (b) The Dirac cone normalised angular frequency of $1.32$ gives a strong directional localisation, as predicted from the analysis of Section \ref{sec03}; the time-harmonic point force generates a localised cross-like waveform. (c) The slight increase in the angular frequency, up to $1.38$, leads to propagating waves along the coordinate axes.}
\label{figmodst4RA2}
\end{figure}

Here we illustrate the uni-directional localization in a homogeneous square lattice made of Rayleigh beams.
The results are presented in Fig.~\ref{figmodst4RA2}.
The flexural displacement field, plotted in these figures, clearly identifies a directional preference for waveforms supported by the square network of the Rayleigh beams.

Three test cases are presented here:
\begin{itemize}
\item The sub-Dirac cone mode, at the normalized  angular frequency of $1.07$; Fig.~\ref{figmodst4RA2}a shows the predicted preferential directions at $45^\circ$ with respect to the $x-$ and $y-$ axes; and it also demonstrates that the evanescent waveforms prevail for this particular vibration mode.
\item The standing wave mode corresponding to the immediate neighbourhood of the Dirac cone, for the normalised angular frequency of $1.32$, where the waveform is highly localised along the coordinate axes aligned with the lattice ligaments, and is shown in Fig.~\ref{figmodst4RA2}b.
\item The propagating directionally localised waveform corresponding to the Dirac cone, above the Dirac vertex frequency, with the value of the normalised angular frequency of $1.38$ is shown in Fig.~\ref{figmodst4RA2}c.
\end{itemize}

\clearpage

\subsection{Neutrality and negative refraction}

The computation presented in this section addresses a square network of beams, which is statically uniform. However, we introduce an interface boundary separating the regions occupied by the Euler-Bernoulli and by the Rayleigh beams, respectively.
A plane flexural wave is incident at the angle of $45^\circ$ from the Euler-Bernoulli region onto the straight interface.

Four illustrative examples are presented below:
\begin{itemize}
\item The long-wave propagation is shown in Fig.~\ref{figmodst4RA3}, for which the Rayleigh and Euler-Bernoulli beams are hardly distinguishable, and hence the composite structure is acting like a homogeneous lattice. The normalised angular frequency is chosen to be $0.06$.
  
\begin{figure}[!htb]
\centering
\includegraphics[width=100mm]{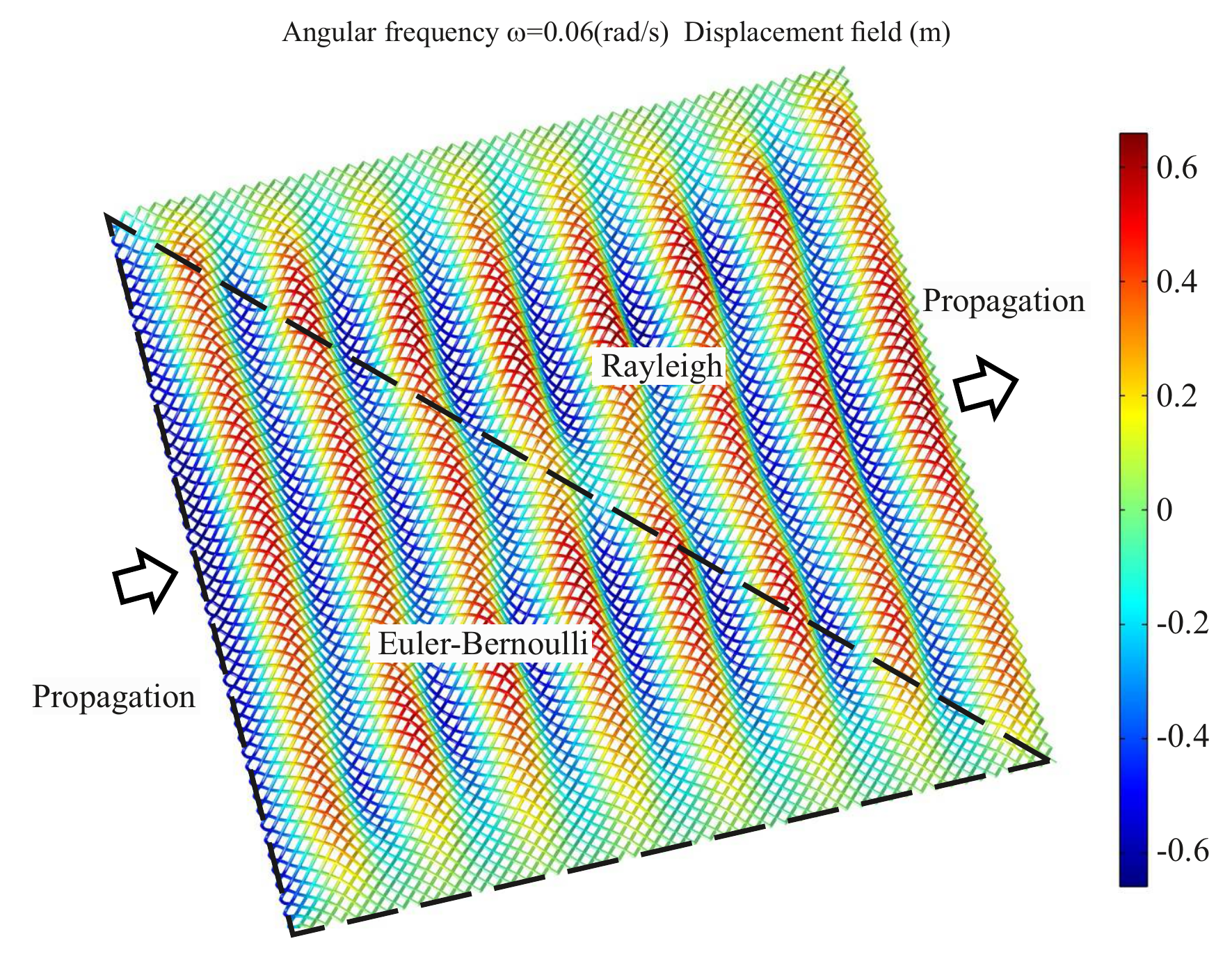}
\caption{\footnotesize
The long-wave illustration for a plane wave propagating across the interface separating the regions occupied by networks of the Euler-Bernoulli and the Rayleigh beams. In this regime the Euler-Bernoulli and Rayleigh beams are hardly distinguishable and the wave propagates as in the homogeneous lattice.}
\label{figmodst4RA3}
\end{figure}

\item Negative refraction has been predicted at the interface between the Euler-Bernoulli and the Rayleigh beams; such example is shown in Fig.~\ref{figmodst4RA4} at the normalised angular frequency of $0.88$, where the incident wave, incoming at the angle of  $45^\circ$, propagates through the lattice of the Rayleigh beams at $-45^\circ$.

\begin{figure}[!htb]
\centering
\includegraphics[width=100mm]{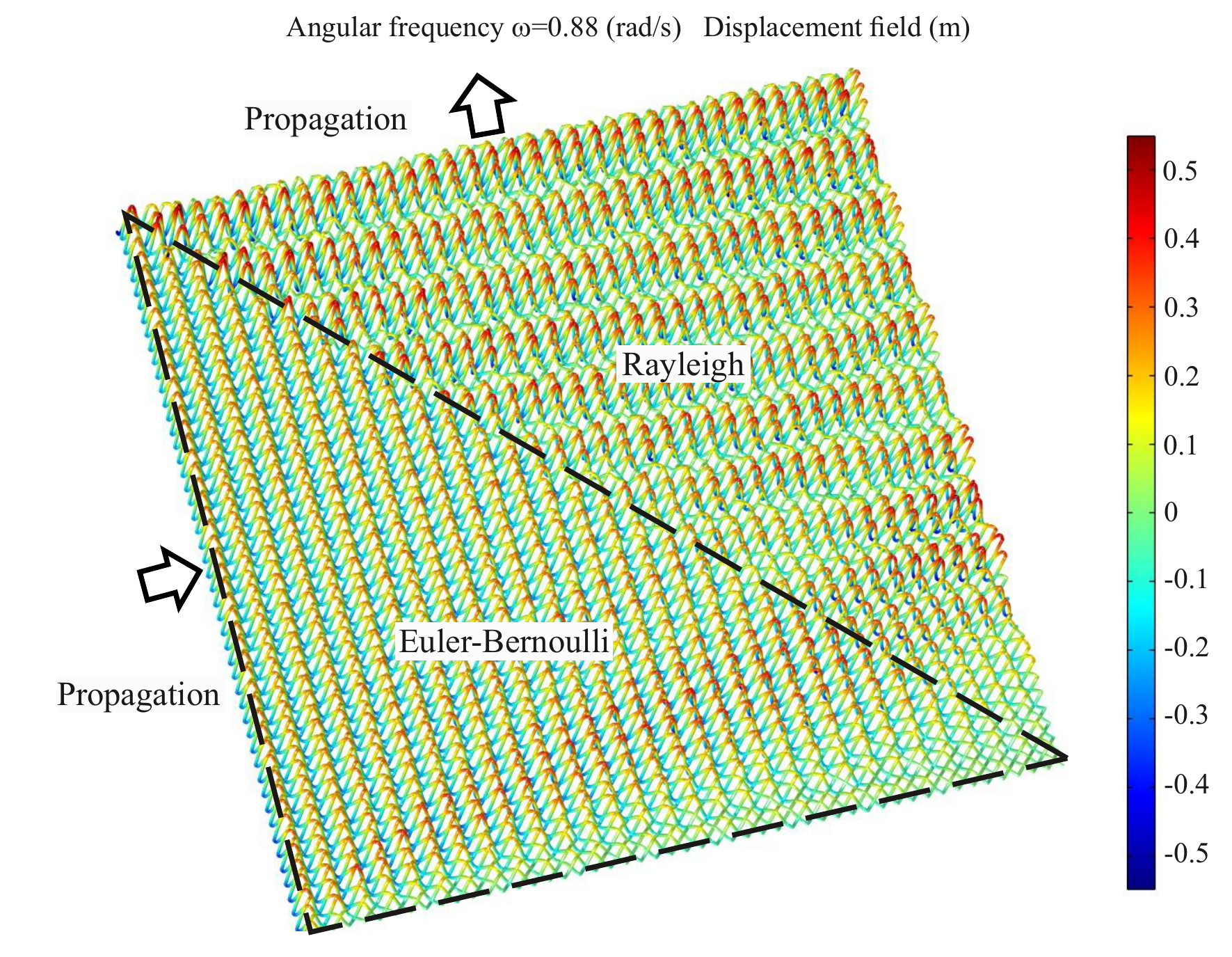}
\caption{\footnotesize  The negative refraction has been predicted for the normalised angular frequency of $0.88$. This is illustrated in the figure, with the directions of propagation of the plane wave shown in arrows.}
\label{figmodst4RA4}
\end{figure}

\item For the sub-Dirac cone mode, at the normalised angular frequency of $1.07$, the evanescent waveforms prevail, which is illustrated in Fig.~\ref{figmodst4RA5}, characterised by the exponentially localised interfacial waveform along the interface.

\begin{figure}[!htb]
\centering
\includegraphics[width=100mm]{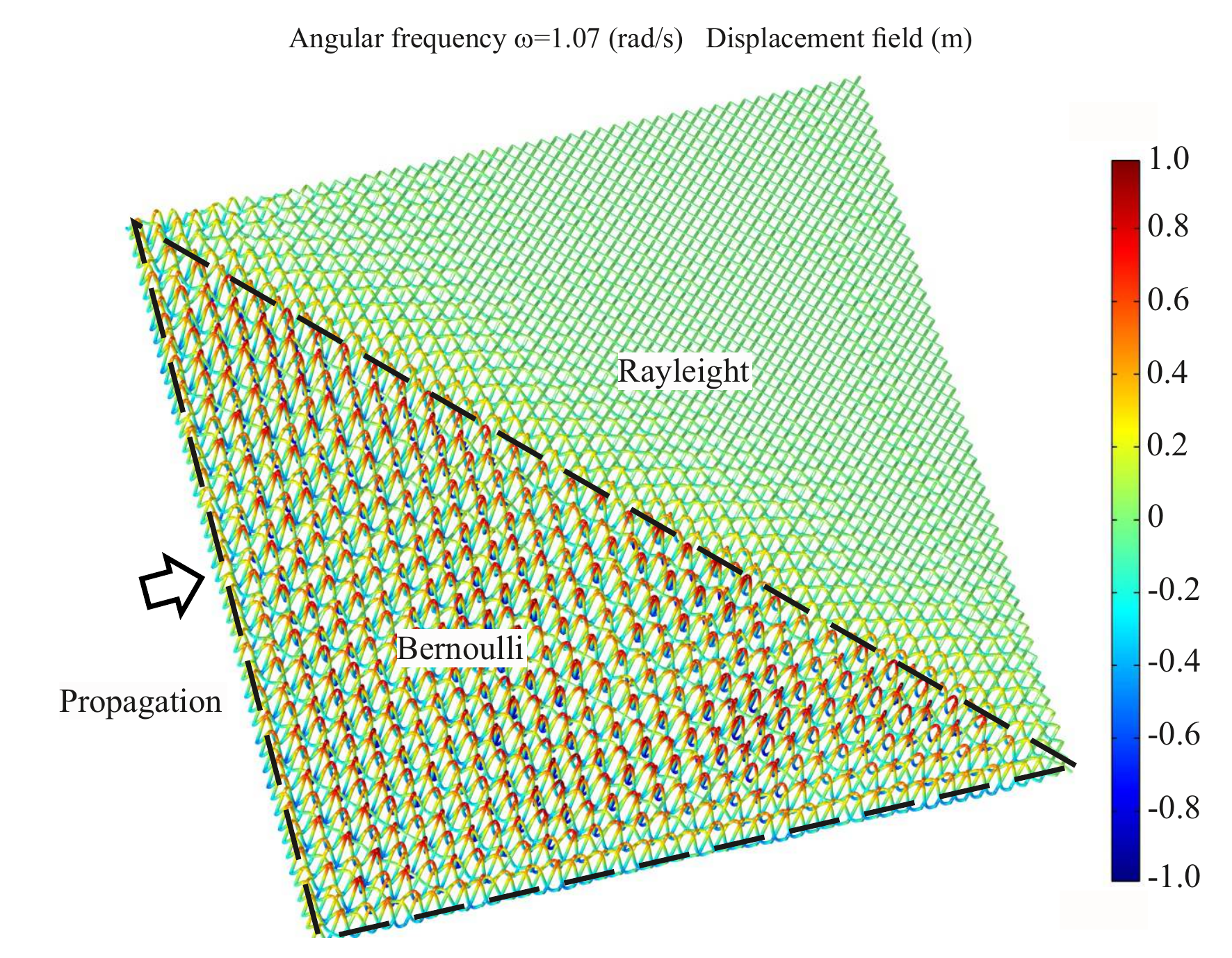}
\caption{\footnotesize Localised waveform at the interface between the networks of Euler-Bernoulli and Rayleigh beams is demonstrated for the case of the normalised angular frequency of $1.07$, which corresponds to a sub-Dirac cone regime.}
\label{figmodst4RA5}
\end{figure}

\item Finally, at the higher normalised angular frequency of $1.76$ on the Dirac cone, above the Dirac cone vertex, a plane wave propagation is expected into the network of the Rayleigh beams, in the same direction as the incident wave.
The prediction is based  on the comparative analysis of the dispersion properties of the flexural waves in the networks of the Euler-Bernoulli and of the Rayleigh beams: at the given frequency the group velocities have the same direction but different magnitudes. Hence the wave front of the modulated wave on the right side from the interface is parallel to the wave front of the incident wave, whilst the larger wavelength of the modulating function is noted.
Indeed, this expectation is fully confirmed and illustrated in Fig.~\ref{figmodst4RA6}.

\begin{figure}[!htb]
\centering
\includegraphics[width=100mm]{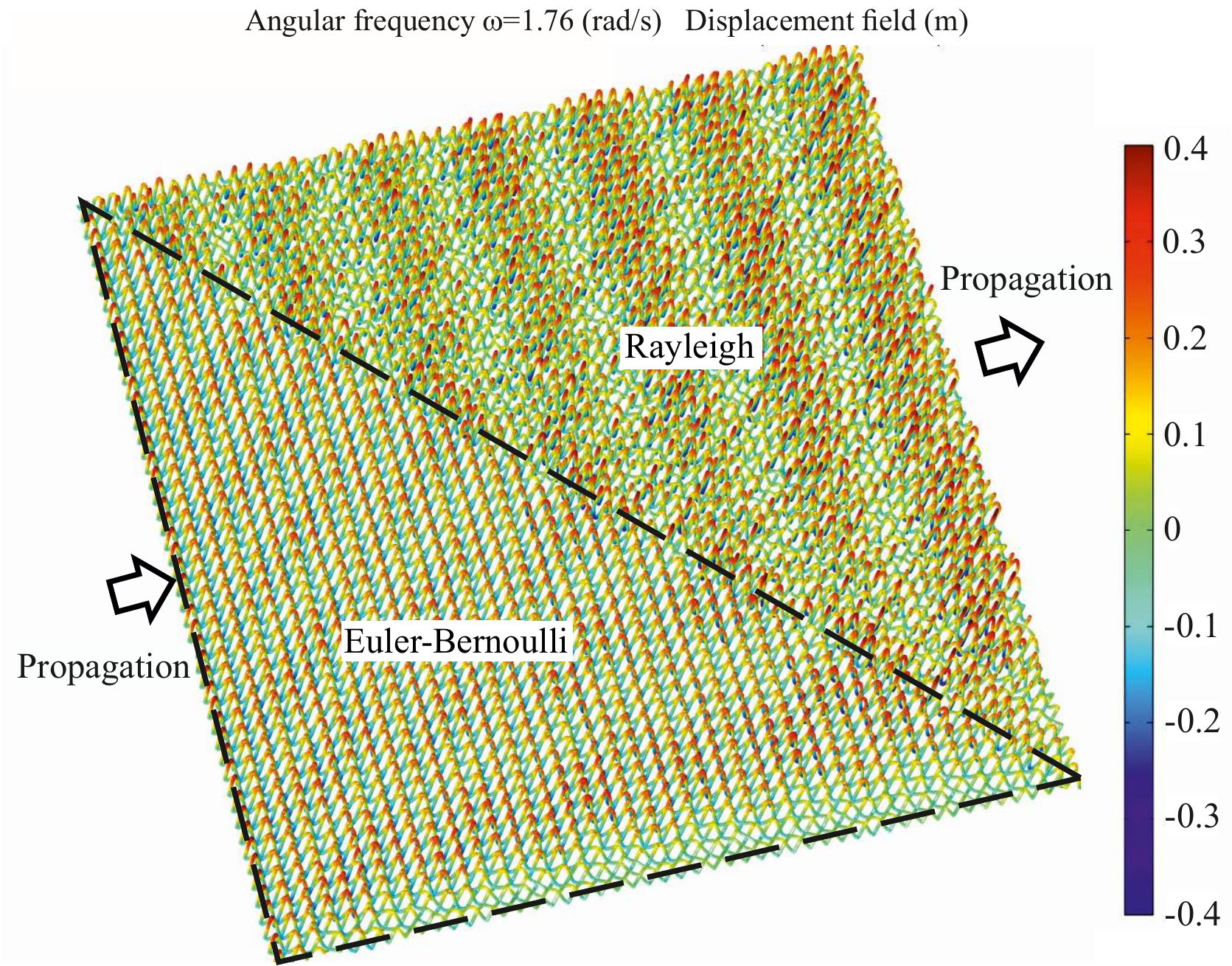}
\caption{\footnotesize
The regime of neutrality gives the plane wave propagating in the same direction as the incident wave. The transmitted plane wave has been modulated according to the features of the Rayleigh beams at the higher normalised angular frequency of $1.76$. }
\label{figmodst4RA6}
\end{figure}

\end{itemize}


\section{Concluding remarks}
\label{sec05}

Dirac points are the points of degeneracies, where the dispersion surface becomes non-smooth, and these points represent the multiple roots of the dispersion equation.
In particular, in our examples the Dirac point represents the root of order $3$.
It is expected that the corresponding Floquet-Bloch waves show a ``neutral plane wave'' propagating through the structure, as well as plane waves of different kinds.
As expected, we see in Figs.~\ref{figmodst3RA} and \ref{figmodst4RA} the Floquet-Bloch waves aligned with the coordinate axes as well as mirror-symmetric waveform modulated along the axes at $\pi/4$ relative to the coordinate axes.
Due to the rotational inertia the frequency of the Dirac point for the lattice consisting of the Rayleigh beams is lower than the frequency of the Dirac point identified for the Euler-Bernoulli beam lattice.

In the above computations, the presence of Dirac cones corresponds to a high level of degeneracy, that occurs for a square lattice of flexural beams.
Any change of symmetry inevitably affects the Dirac cones structure, as the multiple roots of the dispersion equation evolve when a square lattice is replaced by a rectangular lattice, for example.
Dispersion surfaces change dramatically if a square lattice is replaced by a rectangular lattice.

An important feature observed in both cases is the presence of a locally parabolic surface connecting two Dirac cones. This corresponds to so-called ``parabolic metamaterials'' and for the case of continuous flexural plates was observed in \cite{McPhedran_2015}.
Such a case delivers a special dynamic anisotropy of star-shaped waves which were also observed for the so-called hyperbolic metamaterials. However, it is typical that a metamaterial in the parabolic regime would show just one preferential direction instead of two preferential directions observed for hyperbolic metamaterials at frequencies near the saddle points on the dispersion surfaces.

The locally parabolic regimes have been enhanced for waves in the rectangular lattice, compared to the case of a square lattice, and as expected, for the case of an additional
rotational inertia (the Rayleigh beam structure) the dispersion surface move down in the frequency range compared to similar surfaces for the Euler-Bernoulli case.

It is also shown that the Dirac cone observed in the Brillouin zone for the Floquet-Bloch waves in a square lattice, has been split for the case of the rectangular lattice into two Dirac cones, with their vertices being shifted to the boundary of the Brillouin zone. An additional feature of the rectangular lattice simulation is the presence of a ``Dirac edge'' that connects two double roots of the dispersion equation at a lower frequency.

\vspace{6mm}
{\bf Acknowledgements}. AP would like to acknowledge financial support from the
European Union's Seventh Framework Programme FP7/2007-2013/ under REA grant
agreement number PCIG13-GA-2013-618375-MeMic. We acknowledge the support of the research visit of AP in 2015 to the University of Liverpool through the European Union's Grant  PIAPP-GA-284544-PARM-2.
AM has visited the University of Trento in 2016 with the support from the
European Union's Grant ERC-2013-ADG-340561-INSTABILITIES, which is gratefully acknowledged.
AM also acknowledges support from the UK EPSRC Program Grant  EP/L024926/1.
LC acknowledges financial support from the University of Trento, within the research project 2014 entitled ``3D printed metallic foams for biomedical applications: understanding and improving their mechanical behavior''.

\bibliographystyle{jabbrv_unsrt}
\bibliography
{%
roaz1}

\begin{thebibliography}{10}

\bibitem{Slepyan_2002}
L.I. Slepyan.
\newblock {\em Models and Phenomena in Fracture Mechanics}.
\newblock Springer-Verlag, Berlin, 2002.

\bibitem{Slepyan_Ryvkin_2010}
M.~Ryvkin and L.I. Slepyan.
\newblock Crack in a 2d beam lattice: Analytical solutions for two bending
  modes.
\newblock {\em\JournalTitle{Journal of The Mechanics and Physics of Solids}},
  58, No6:902--917; DOI: 10.1016/j.jmps.2010.03.006, 2010.

\bibitem{Heckl_2002}
M.A. Heckl.
\newblock Coupled waves on a periodically supported timoshenko beam.
\newblock {\em\JournalTitle{Journal of Sound and Vibration}}, 252(5):849--882,
  2002.

\bibitem{Brun_2013}
M.~Brun, A.B. Movchan, and L.I. Slepyan.
\newblock Transition wave in a supported heavy beam.
\newblock {\em\JournalTitle{J. Mech. Phys. Solids}}, 61 (10):2067--2085, 2013.

\bibitem{Bigoni_2002}
D.~Bigoni and A.B. Movchan.
\newblock Statics and dynamics of structural interfaces in elasticity.
\newblock {\em\JournalTitle{International Journal of Solids and Structures}},
  39(19):4843--4865, 2002.

\bibitem{Gei2009}
M.~Gei, A.B. Movchan, and D.~Bigoni.
\newblock Band-gap shift and defect-induced annihilation in prestressed elastic
  structures.
\newblock {\em\JournalTitle{Journal of Applied Physics}}, 105(6), 2009.

\bibitem{PM_2014}
A.~Piccolroaz and A.B. Movchan.
\newblock Dispersion and localization in structured rayleigh beams.
\newblock {\em\JournalTitle{International Journal of Solids and Structures}},
  51:4452--4461, 2014.

\bibitem{Timoshenko1990}
W.~Weaver Jr., S.~P. Timoshenko, and D.~H. Young.
\newblock {\em Vibration Problems in Engineering}.
\newblock John Wiley \& Sons, New York, 1990.

\bibitem{HAN1999935}
S.M. Han, H.~Benaroya, and T.~Wei.
\newblock Dynamics of transversely vibrating beams using four engineering
  theories.
\newblock {\em\JournalTitle{Journal of Sound and Vibration}}, 225(5):935 --
  988, 1999.

\bibitem{Li20111677}
X.-F. Li, Z.-W. Yu, and H.~Zhang.
\newblock Free vibration of shear beams with finite rotational inertia.
\newblock {\em\JournalTitle{Journal of Constructional Steel Research}},
  67(10):1677 -- 1683, 2011.

\bibitem{BLP1978}
A.~Bensoussan, J.L. Lions, and G.~Papanicolaou.
\newblock {\em Asymptotic analysis for periodic structures}.
\newblock North-Holland, Amsterdam, 1978.

\bibitem{SP1980}
E.~Sanchez-Palencia.
\newblock {\em Non homogeneous media and vibration theory. Lecture Notes in
  Physics 127}.
\newblock Springer-Verlag, Berlin, 1980.

\bibitem{Jikov1994}
V.V. Jikov, S.M. Kozlov, and O.~A. Oleinik.
\newblock {\em Homogenization of Differential Operators and Integral
  Functionals}.
\newblock Springer-Verlag, Berlin, 1994.

\bibitem{Physics_2013}
M.~Brun, A.B. Movchan, I.S. Jones, and R.C. McPhedran.
\newblock Bypassing shake, rattle and roll.
\newblock {\em\JournalTitle{Physics World}}, 26 (5):32--36, 2013.

\bibitem{Craster_2014}
T.~Antonakakis, R.V. Craster, and S.~Guenneau.
\newblock Homogenisation for elastic photonic crystals and dynamic anisotropy.
\newblock {\em\JournalTitle{Journal of the Mechanics and Physics of Solids}},
  71:84--96, 2014.

\bibitem{McPhedran_2015}
R.C. McPhedran, A.B. Movchan, N.V. Movchan, M.~Brun, and M.J.A. Smith.
\newblock 'parabolic' trapped modes and steered dirac cones in platonic
  crystals.
\newblock {\em\JournalTitle{Proc. R. Soc. Lond. A}}, 471:20140746, 2015.

\bibitem{Panasenko2005}
G.P. Panasenko.
\newblock {\em Multi-Scale Modelling for Structures and Composites}.
\newblock Springer-Verlag, Berlin, 2005.

\bibitem{Fleck2006}
A.~Srikantha~Phani, J.~Woodhouse, and N.A. Fleck.
\newblock Wave propagation in two-dimensional periodic lattices.
\newblock {\em\JournalTitle{J. Acoust. Soc. Am.}}, 119 (4):1995--2005, 2006.

\end{thebibliography}

%
%

\end{document}